\documentclass[12pt,a4paper]{article}
\usepackage[textheight=23cm,textwidth=17.5cm]{geometry}
\usepackage{amsmath}
\usepackage{graphicx}
\usepackage{pdfpages}
\usepackage{cite}
\usepackage{here}
\usepackage[american]{babel}
\usepackage{amssymb}
\usepackage{verbatim}
\usepackage{textcomp}
\usepackage{threeparttable}

\usepackage{dcolumn} 
\newcolumntype{d}{D{.}{.}{2}}
\newcolumntype{e}{D{.}{.}{3}}
\newcolumntype{f}{D{.}{.}{4}}
\newcommand{\fett}[1]{\mbox{\boldmath$#1$}}

\newcommand{\bra}[1]{\ensuremath{\left\langle #1\right|}}
\newcommand{\ket}[1]{\ensuremath{\left|#1\right\rangle}}
\newcommand{\braket}[2]{\ensuremath{\left\langle #1\vphantom{#2}\right.\left|\vphantom{#1}#2\right\rangle}}
\newcommand{\braOket}[3]{\ensuremath{\left\langle #1\vphantom{#2#3}\right.\left|\vphantom{#1#3}#2\right|\left.\vphantom{#1#2}#3\right\rangle}}

\DeclareMathAlphabet{\mathpzc}{OT1}{pzc}{m}{it}
\def\trs{^\mr{T}}
\newcommand{\mx}[1]{{\boldsymbol{#1}}}
\newcommand{\mr}[1]{\mathrm{#1}}
\def\Eh{E$_\mr{h}$}
\usepackage{color}
\usepackage{amsmath}

\begin{document}

\begin{center}
~\vspace*{-0.01cm}
{\Large %
 Electric Transition Dipole Moment in pre-Born--Oppenheimer Molecular Structure Theory %
}\\
\vspace{0.3cm}
{\large
Benjamin Simmen$^a$, Edit M\'atyus$^b$\footnote{corresponding author; e-mail: matyus@chem.elte.hu}, Markus Reiher$^a$\footnote{corresponding author; e-mail: markus.reiher@phys.chem.ethz.ch}
}\\[2ex]

$^a$ETH Z\"urich, Laboratorium f\"ur Physikalische Chemie, \\
Vladimir-Prelog-Weg 2, 8093 Z\"urich, Switzerland\\
$^b$E\"ovt\"os University, Institute of Chemistry, P.O. Box 32, H-1518, Budapest 112, Hungary\\Present Address: Department of Chemistry, University of Cambridge, Lensfield Road, Cambridge, CB2 1EW, United Kingdom\\[3ex]
14.06.2014\\[-0.5cm]

\end{center}

\abstract This paper presents the calculation of the electric transition dipole moment in a pre-Born--Oppenheimer framework. Electrons and nuclei are treated equally in terms of the parametrization of the non-relativistic total wave function, which is written as a linear combination of basis functions constructed with explicitly correlated Gaussian functions and the global vector representation. The integrals of the electric transition dipole moment are derived  corresponding to these basis functions in both the length and the velocity representation. The complete derivation and the calculations are performed in laboratory-fixed Cartesian coordinates without relying on coordinates which separate the center of mass from the translationally invariant degrees of freedom. The effect of the overall motion is eliminated via translationally invariant integral expressions. As a numerical example the electric transition dipole moment is calculated between two rovibronic levels of the H$_2$ molecule assignable to the lowest rovibrational states of the $X\ ^1\Sigma^+_\mr{g}$ and $B\ ^1\Sigma^+_\mr{u}$ electronic states in the clamped-nuclei framework. This is the first evaluation of this quantity in a full quantum mechanical treatment without relying on the Born--Oppenheimer approximation.
\newpage
\normalsize
\setlength{\parindent}{0cm}
\setlength{\parskip}{2ex}

\section{Introduction}

In this work, we are interested in the evaluation of the electric transition dipole moment. In a semi-classical picture, it mediates transitions between two rovibrational or rovibronic states of a molecule induced by the electric component of the classical electromagnetic radiation field. 

Almost all calculations presently carried out on molecular systems rely on the Born--Oppenheimer (BO) approximation, which separates the electronic and nuclear degrees of freedom. For small systems, however, very accurate calculations can be carried out if the BO approximation is avoided. The simultaneous description of electrons and nuclei using explicitly correlated Gaussian functions has been pioneered by the Adamowicz group \cite{Bubin2012,Mitroy2013} and by Suzuki and Varga \cite{Varga1998,Suzuki1998,Suzuki1998a}. 

Here, we follow these lines and extend our earlier work \cite{Matyus2012,Simmen2013,Matyus2013}. The basis functions are constructed from explicitly correlated Gaussian functions \cite{Jeziorski1979,Cencek1993,Rychlewski2004,Boys1960,Singer1960} and the global vector representation \cite{Varga1998,Suzuki1998,Suzuki1998a} in order to ensure that the wave function is an eigenfunction of the total spatial angular momentum operators, $\hat{L}^2$ and $\hat{L}_z$, and parity. The parameters of the basis functions are optimized variationally through stochastic sampling. 

Rather than relying on a set of Cartesian coordinates which separates the translationally invariant Cartesian coordinates (TICC) from the center of mass, we use laboratory-fixed Cartesian coordinates. Any translational contamination to the total energy is eliminated from the integrals as presented in our recent paper\cite{Simmen2013}. We illustrate in this work that this scheme can be applied to the calculation of molecular properties such as the electric transition dipole moment. Previous calculations in the literature of the electric transition dipole moment in a pre-BO framework have been performed in TICC\cite{Tian2012,Bekbaev2011}.

In Section \ref{sec:theory}, we present the pre-BO framework used in this work and the most essential conceptual aspects of the electric permanent and transition dipole moments. In Section \ref{sec:dipole}, we derive the translationally invariant integral expressions of the electric transition dipole moment components. In Section \ref{sec:results}, we discuss numerical results for transitions of the four-particle H$_2=\lbrace\mr{p}^+,\mr{p}^+,\mr{e}^-,\mr{e}^-\rbrace$ molecule.

\section{Theoretical Background}
\label{sec:theory}
In this section, we present the theoretical background relevant to this work. We begin with a short introduction to the variational procedure of pre-BO molecular structure theory based on explicitly correlated Gaussian functions (ECGs) \cite{Jeziorski1979,Cencek1993,Rychlewski2004,Boys1960,Singer1960} in the global vector representation (GVR) \cite{Varga1998,Suzuki1998,Suzuki1998a}. Furthermore, we discuss some fundamental aspects of the electric transition and permanent dipole moment in the pre-BO theory.

\subsection{pre-Born--Oppenheimer Molecular Structure Theory}

The non-relativistic quantum Hamiltonian of an $(n+1)$-particle system of nuclei and electrons in atomic units is
\begin{align}
  \hat{H}
  =
  -\sum_{i=1}^{n+1} \frac{1}{2m_i} \Delta_{\mx{r}_i}
  +
  \sum_{i=1}^{n+1}\sum_{j>i}^{n+1} \frac{q_iq_j}{|\mx{r}_i-\mx{r}_j|}
  \label{eq:Hop}
\end{align}
where $\fett{r}^{\mr{T}}=(\fett{r}_1^{\mr{T}},\ldots,\fett{r}_{n+1}^{\mr{T}})$ is the collection of the laboratory-fixed Cartesian coordinates (LFCC) and $m_i$ is the mass and $q_i$ is the electric charge of the $i^{th}$ particle. The time-independent Schr\"odinger equation for this Hamiltonian is solved variationally by writing the wave function as a linear combination of basis functions, which also contain (non-linear) variational parameters.

The spatial part of the basis functions used in this work is constructed from explicitly correlated Gaussian functions (ECG) and the global vector representation (GVR). The basis functions are eigenfunctions of the total spatial angular momentum operators, denoted by $\hat{L}^2$ and $\hat{L}_z$, with the corresponding quantum numbers, $L$ and $M_L$, as well as the space inversion operator with parity $p=(-1)^L$ (only natural-parity states are considered in this work).  Then, a spatial basis function is\cite{Suzuki1998a}
\begin{align}
  \phi_{LM_L}(\mx{r};\mx{A},\mx{u},K)
  &=
  |\mx{v}|^{2K+L}
  Y^{L}_{M_{L}}(\hat{\mx{v}}) \exp\left(-\frac{1}{2} \mx{r}\trs (\mx{A}\otimes \mx{1}_3) \mx{r} \right)
  \label{eq:basdef2}
\end{align}
with the global vector
\begin{gather}
  \mx{v} = (\mx{u}\otimes \mx{1}_3) \mx{r} \ .
\end{gather}
In Eq.\ (\ref{eq:basdef2}), $\fett{A}$, $\fett{u}$ and $K$ are variational parameters. The matrix $\mx{A}$ describes the pairwise correlations among the elements of $\fett{r}$. It is positive-definite in order to obtain basis functions which are square-integrable and have a non-vanishing norm. Further variational parameters are the vector $\mx{u}=(u_1,\ldots,u_{n+1})$ determining the direction and magnitude of the so-called global vector and the exponent $K$ is a non-negative integer.

The spherical harmonic function $Y^{L}_{M_{L}}$ of degree $L$ and order $M_L$ describes the angular part of the quantum system. The Condon--Shortley phase convention is used for $Y^{L}_{M_{L}}$ throughout this work. We note that the quantum number $L$ corresponds here to the total spatial angular momentum operator (the vector sum of the orbital and rotational angular momentum), which is a common notation in the literature of quantum mechanical few-particle systems (see for example Refs. \cite{Mitroy2013,Varga1998,Suzuki1998,Suzuki1998a}, Refs. \cite{Tian2012,Bekbaev2011}). Hence, this $L$ is equivalent to $N$ often used in molecular spectroscopy according to the IUPAC recommendations \cite{Ion}.

The basis function described in Eq.\ (\ref{eq:basdef2}) can also be formulated in terms of a generating function\cite{Suzuki1998a}:
\begin{gather}
\phi_{LM_L}(\mx{r};\mx{A},\mx{u},K)=\frac{1}{B_{KL}}\int d\hat{\fett{\varepsilon}}  Y^{L}_{M_{L}}(\hat{\fett{\varepsilon}})\left\{\frac{\partial^{2K+L}}{\partial a^{2K+L}} g(\fett{r};\mx{A},a\mx{u}\otimes\fett{\varepsilon})\right\}_{a=0,|\fett{\varepsilon}|=1}
\end{gather}
with the generating function
\begin{gather}
 g(\fett{r};\mx{A},a\mx{u}\otimes\fett{\varepsilon})=\exp\left(-\frac{1}{2}\fett{r}\trs(\mx{A}\otimes\fett{1}_3)\fett{r}+(a\mx{u}\otimes\fett{\varepsilon})\trs \fett{r})\right) \label{eq:genfun}
\end{gather}
and
\begin{gather}
 B_{mL}=\frac{4\pi(L+2m)!(L+m+1)!2^{L+1}}{m!(2L+2m+2)!}\ .
\end{gather}
We have found this form is particularly convenient for the derivation of integral expressions.

The total basis functions also include a spin part, constructed from the eigenfunctions of the elementary spin operators ($\hat{S}^2_\mr{e}$, $\hat{S}_{z\mr{e}}$) and ($\hat{S}^2_\mr{p}$, $\hat{S}_{z\mr{p}}$) of the electrons and protons\cite{Matyus2012}. The particle-exchange symmetry is imposed on the total basis functions through an explicit anti-symmetrization procedure. For more details, the reader is referred to Ref.\ \cite{Matyus2012}. The non-linear parameters of the basis functions are partially taken from earlier calculations, where the parameter selection and optimization have been described in detail \cite{Matyus2012}. Parameters generated for this work are obtained accordingly.

The generalized eigenvalue problem corresponding to the matrix representation of the Hamiltonian
is solved by the standard linear algebra library routines of LAPACK (Version 3.2.1)\cite{Anderson1999} through the Armadillo framework (Version 3.4.0)\cite{Sanderson2010}.

\subsection{Electric Permanent and Transition Dipole Moments in Pre-BO Theory}

As we are interested in the components of the electric transition dipole moment in pre-BO theory. In systems containing $n+1$ particles the electric dipole moment operator is defined as
\begin{gather}
 \hat{\fett{\mu}}=\sum_{i=1}^{n+1}q_i\fett{r}_i\label{eq:dipolemoment}
\end{gather}
where $q_i$ and $\fett{r}_i$ are the electric charge and position of particle $i$, respectively. This operator has odd parity and describes both permanent and transition dipole moments. But although being closely related, there are significant conceptual differences between the two types of dipole moments which we will discuss in this section.

Molecules described in the BO approximation have generally a non-zero permanent electric dipole moment (i.e.\ they are polar). The only exceptions are molecules which are non-polar due to certain symmetry properties contained within the molecular structure (e.g.\ methane, benzene or H$_2$). This illustrates how closely the molecular structure and the permanent electric dipole are related. The concept of structure (or shape) where the nuclei form a rigid scaffold, which is stabilized by the electrons, is the core concept in chemistry and clearly defined in the BO picture. Yet, there is currently no complete understanding how this concept is to be interpreted in a pre-BO framework since the nuclei are not fixed but treated as quantum particles to which particle densities are assigned. Several authors  have been discussing the subject in great detail\cite{Matyus2011a,Matyus2011,Rodriguez2013,Muller-Herold2006,Muller-Herold2008,Ludena2012,Becerra2013,Goli2012,Goli2013,Aguirre2013,King2013,Perez-Torres2013,Chakraborty2008,Sutcliffe2005a,Sutcliffe2005b,Woolley1976,Woolley1998,Woolley1991}.

Generally, the symmetry properties of a pre-BO wave function are enough to gain some information about pre-BO permanent dipole moments. The total  pre-BO wave function is an eigenfunction of the parity operator if no external potential is present. This is due to the isotropy of space and the resulting conservation of the total spatial angular momentum. The parity of the pre-BO wave function together with the odd parity of the dipole moment operator results in an integral over a function with ungerade symmetry and therefore always evaluates to zero. Yet, the squared length of the dipole moment
\begin{gather}
 \hat{\fett{\mu}}\trs\hat{\fett{\mu}}=\sum_{j=1}^{n+1}\sum_{i=1}^{n+1}q_iq_j\fett{r}\trs_i\fett{r}_j\ ,\label{eq:dipolelength}
\end{gather}
has even parity. Thus, the integral of the squared length has gerade symmetry and is not strictly zero. We can conclude from this that a pre-BO wave function has no permanent dipole moment but can still be polarized. This indicates that the  permanent dipole vanishes because it has no preferred orientation due to the symmetry properties of the wave function. This idea might be investigated in later work. Another method to determine the permanent transition dipole moment (and higher-order electric properties) was presented by Cafiero, Bubin and Adamowicz\cite{Cafiero2003}. They included the interaction energy of an external electric field with the permanent electric dipole in the total Hamiltonian and extrapolated the zero field energy and electric properties from results at various field strengths.

In contrast to the status of the permanent dipole moment, which is related to the classical chemical structure concept, the evaluation of the electric transition dipole moment, which is a spectroscopic quantity, is more straightforward in the pre-BO theory. The pre-BO wave function already contains not only the electronic but also the rotational-vibrational degrees of freedom.

The subscripts $\mr{i}$ and $\mr{f}$ denote the initial and final states of a transition. Transitions mediated through an electric transition dipole feature a set of selection rules such that
\begin{gather}
 \fett{\mu}_{\mr{if}}=\bra{\Psi_\mr{i}}\hat{\fett{\mu}}\ket{\Psi_\mr{f}}\neq0\ .\label{eq:selectionrule}
\end{gather}
Electric transition dipole moments are complex three-vectors of the form $\fett{\mu}=(\mu_{x},\mu_{y},\mu_{z})^{\rm{T}}$. The length of the electric transition dipole moment is then
\begin{gather}
|\fett{\mu}|=(\fett{\mu}^\dag\fett{\mu})^{1/2}\ .
\end{gather}

For the spatial angular quantum numbers $(L_\mr{i},M_{L\mr{i}})$ and $(L_\mr{f},M_{L\mr{f}})$, with natural parity $p_\mr{i}=(-1)^{L_\mr{i}}$ and $p_\mr{f}=(-1)^{L_\mr{f}}$, the integral in Eq.\ (\ref{eq:selectionrule}) is non-vanishing if $L_\mr{i}-L_\mr{f}\in\{+1,-1\}$ and for $M_{L\mr{i}}-M_{L\mr{f}}\in\{+1,0,-1\}$. The selection rules for $M_L$ are related to different polarizations of the absorbed/emitted light. For the spin quantum numbers one finds $S_\mr{i}-S_\mr{f}=M_{S\mr{i}}-M_{S\mr{f}}=0$. Transitions which do not fulfill these selection rules are either symmetry forbidden (selection rules related to $L$ and $M_L$) or spin forbidden (selection rules related to $S$ and $M_S$). Spin forbidden transitions become allowed if different spin states mix, e.g, if relativistic effects are considered. Symmetry forbidden transitions become allowed with respect to transitions mediated by means of higher-order electric transition multi-poles.

Furthermore, degenerate substates have to be considered: Transitions involve all rotational substates $\Psi_{L,M_L}$ with $-L\leq M_L\leq L$ and all possible transitions for which Eq.\ (\ref{eq:selectionrule}) is fulfilled. The squared length of the transition dipole moment is then obtained as\cite{Hilborn1982}
\begin{gather}
\left|\braOket{\Psi_{L_\mr{i}}}{\hat{\fett{\mu}}}{\Psi_{L_\mr{f}}}\right|^2=\sum_{j=-L_\mr{i}}^{L_\mr{i}}\sum_{k=-L_\mr{f}}^{L_\mr{f}}\left|\braOket{\Psi_{L_\mr{i},j}}{\hat{\fett{\mu}}}{\Psi_{L_\mr{f},k}}\right|^2\ .\label{eq:mu}
\end{gather}

\section{Evaluation of the Electric Transition Dipole Moment Integrals}
\label{sec:dipole}
In this section, we focus on the electric transition dipole moment and its determination from explicitly correlated Gaussian functions with the global vector representation in an laboratory-fixed Cartesian coordinate pre-BO framework. First, we recall two different forms of the electric transition dipole moment, commonly known as the velocity and the length representation. The two representations are equivalent only for the exact wave function. Then, expressions are presented for the transition dipole integrals.

\subsection{Velocity and Length Representation}

The transition dipole moment between the ``$\mr{i}$'' initial and ``$\mr{f}$'' final state is
\begin{gather}
 \fett{\mu}^\mr{(l)}_\mr{if}=\sum_{j=1}^{n+1}\braOket{\Psi_{\mr{i}}}{q_j\fett{r}_j}{\Psi_{\mr{f}}}\label{eq:length}
\end{gather}
in the ``length'' ($\mr{l}$) representation, while it can also be evaluated in the ``velocity'' ($\mr{v}$) representation \cite{Chandrasekhar1945}
\begin{gather}
 \fett{\mu}^\mr{(v)}_\mr{if}=-\frac{1}{(E_\mr{i}-E_\mr{f})}\sum_{j=1}^{n+1}\braOket{\Psi_{\mr{i}}}{\frac{q_j}{m_j}\fett{\nabla}_{\fett{r}_j}}{\Psi_{\mr{f}}}\ .\label{eq:velocity}
\end{gather}
 The equivalence of Eqs. (\ref{eq:length}) and (\ref{eq:velocity}) can be shown with the help of the commutation relation
\begin{gather}
 [\hat{H},q_j\fett{r}_j]=-\frac{q_j}{m_j}\fett{\nabla}_{\fett{r}_j}\ .
\end{gather}
Acting with the initial state from the left and the final state from the right leads to the integral
\begin{gather}
 \bra{\Psi_\mr{i}}[\hat{H},q_j\fett{r}_j]\ket{\Psi_\mr{f}}=-\bra{\Psi_\mr{i}}\frac{q_j}{m_j}\fett{\nabla}_{\fett{r}_j}\ket{\Psi_\mr{f}}\ .
\end{gather}
Exploiting the fact, that the inital and the final state are eigenfunction of the Hamiltonian we find
\begin{gather}
 (E_\mr{i}-E_\mr{f})\bra{\Psi_\mr{i}}q_j\fett{r}_j\ket{\Psi_\mr{f}}=-\bra{\Psi_\mr{i}}\frac{q_j}{m_j}\fett{\nabla}_{\fett{r}_j}\ket{\Psi_\mr{f}}\label{eq:equivalence}
\end{gather}
where $\mr{E}_\mr{i}$ and $\mr{E}_\mr{f}$ are the energies of the initial and the final states, respectively. Eq.\ (\ref{eq:equivalence}) can be easily rearranged into
\begin{gather}
 \bra{\Psi_\mr{i}}q_j\fett{r}_j\ket{\Psi_\mr{f}}=-\frac{1}{(\mr{E}_\mr{i}-\mr{E}_\mr{f})}\bra{\Psi_\mr{i}}\frac{q_j}{m_j}\fett{\nabla}_{\fett{r}_j}\ket{\Psi_\mr{f}} \label{eq:equifinal}
\end{gather}
which are the individual terms of the sums in Eqs.\ (\ref{eq:length}) and (\ref{eq:velocity}). The equality in Eq.\ (\ref{eq:equifinal}) holds only for exact wave functions. At the same time, no conclusion can be drawn on which representation is less sensitive to approximations in the wave functions \cite{Anderson1974}.

\subsection{Evaluation of the Integrals}
The following three-step evaluation procedure used already for the evaluation of the overlap, kinetic and Coulomb potential energy integrals \cite{Matyus2012} is employed in the present work for the calculation of the transition dipole moment integrals in both the length and velocity representation. The three steps for some operator $\hat{O}$ are:
\begin{enumerate}
 \item Evaluate the integrals of $\hat{O}$ with the generating functions of Eq.\ (\ref{eq:genfun}):
 \begin{gather}
  I_{1}=\bra{g(\fett{r},\fett{A}_I,a_I\mx{u}_I\otimes\fett{\varepsilon}_I)}\hat{O}\ket{g(\fett{r},\fett{A}_J,a_J\mx{u}_J\otimes\fett{\varepsilon}_J)}\ .
 \end{gather}
 \item Evaluate the derivatives at $a_I=a_J=0$:
\begin{gather}
  I_{2}=\frac{\partial^{2K_I+L_I}}{\partial a_I^{2K_I+L_I}}\frac{\partial^{2K_J+L_J}}{\partial a_J^{2K_J+L_J}}I_{1}(a_I,a_J)\bigg|_{a_I=a_J=0}\ .
\end{gather}
 \item Evaluate the angular integrals:
\begin{gather}
I_3=\frac{1}{B_{K_IL_I}B_{K_JL_J}}\int d \hat{\fett{\varepsilon}}_I\int d \hat{\fett{\varepsilon}}_J Y^{L_I\ast}_{M_{LI}}(\hat{\fett{\varepsilon}}_I)Y^{L_J}_{M_{LJ}}(\hat{\fett{\varepsilon}}_J)I_{2}(\fett{\varepsilon}_I,\fett{\varepsilon}_J)\ .
\end{gather}

\end{enumerate}
In order to avoid numerical instabilities, quasi-normalized basis functions are used, i.e., each basis function is normalized with respect to its spatial part. The ($I^{th}$,$J^{th}$) matrix elements for operator $\hat O$ are then
\begin{gather}
 [O]_{IJ}=\frac{\bra{\phi_I^{(L_I,M_{LI})}(\mx{r};\mx{A}_I,\mx{u}_I,K_I)}\hat O \ket{\phi_J^{(L_J,M_{LJ})}(\mx{r};\mx{A}_J,\mx{u}_J,K_J)}}{\left|\phi_I^{(L_I,M_{LI})}\right|\left|\phi_J^{(L_J,M_{LJ})}\right|}\ ,
\end{gather}
 where $\left|\phi_I^{(L_I,M_{LI})}\right|$ and $\left|\phi_J^{(L_J,M_{LJ})}\right|$ are the normalization factors and the square brackets denote the matrix representation of $\hat O$.

In this work, the integrals in Eqs.\ (\ref{eq:length}) and (\ref{eq:velocity}) are evaluated with ECG basis functions in the GVR. Instead of the original Cartesian components $\alpha\in\{x,y,z\}$ we use transformed components
\begin{gather}
 (\hat{\fett{\Omega}}_j)_+=(\hat{\fett{\Omega}}_j)_x-i(\hat{\fett{\Omega}}_j)_y\label{eq:omegap}\\
(\hat{\fett{\Omega}}_j)_-=(\hat{\fett{\Omega}}_j)_x+i(\hat{\fett{\Omega}}_j)_y\label{eq:omegam}\\
(\hat{\fett{\Omega}}_j)_z=(\hat{\fett{\Omega}}_j)_z\label{eq:omegaz}
\end{gather}
collected under the label $\beta\in\{+,-,z\}$ where $\hat{\fett{\Omega}}_j\in\{\hat{\fett{\mu}}^\mr{(v)},\hat{\fett{\mu}}^\mr{(l)}\}$. The transformed components are especially convenient for evaluating the angular integrals. For the complete derivation we may refer the reader to the supporting information.

Following this integration scheme we get the expression:
\begin{gather}
  [\fett{\Omega}_\beta]_{IJ}=\frac{\bra{\phi_I^{(L_I,M_{LI})}}\hat{\fett{\Omega}}_\beta\ket{\phi_J^{(L_J,M_{LJ})}}}{\left|\phi_I^{(L_I,M_{LI})}\right|\left|\phi_J^{(L_J,M_{LJ})}\right|}\nonumber\\
 =\left(\frac{\det(2\mx{A}_I)\det(2\mx{A}_J)}{\det(\mx{A}_{IJ})\det(\mx{A}_{IJ})}\right)^{3/4}\left(\frac{p_{II}}{q_I}\right)^{K_I}\left(\frac{p_{JJ}}{q_J}\right)^{K_J}\left(\frac{p_{IJ}}{\sqrt{q_Iq_J}}\right)^{L_J}C^\beta_1(L_J,L_I,M_{LJ},M_{LI})\left(q_I\right)^{-1/2}\nonumber\\
\times\left[G^{\mx{\Omega}}_I\sum_{m=0}^{\min(K_I,K_J)}\left(\frac{p_{IJ}p_{IJ}}{p_{II}p_{JJ}}\right)^m H_1(m,K_I,K_J,L_J)\right.\nonumber\\
\left.+\frac{G^{\mx{\Omega}}_Jp_{II}}{p_{IJ}}\sum_{m=1}^{\min(K_I+1,K_J)}\left(\frac{p_{IJ}p_{IJ}}{p_{II}p_{JJ}}\right)^m H_2(m,K_I,K_J,L_I)\right]\nonumber\\
+\left(\frac{\det(2\mx{A}_I)\det(2\mx{A}_J)}{\det(\mx{A}_{IJ})\det(\mx{A}_{IJ})}\right)^{3/4}\left(\frac{p_{II}}{q_I}\right)^{K_I}\left(\frac{p_{JJ}}{q_J}\right)^{K_J}\left(\frac{p_{IJ}}{\sqrt{q_Iq_J}}\right)^{L_I}C^\beta_2(L_I,L_J,M_{LI},M_{LJ})\left(q_J\right)^{-1/2}\nonumber\\
\times\left[\frac{G^{\mx{\Omega}}_Ip_{JJ}}{p_{IJ}}\sum_{m=1}^{\min(K_I,K_J+1)}\left(\frac{p_{IJ}p_{IJ}}{p_{II}p_{JJ}}\right)^m H_2(m,K_J,K_I,L_J)\right.\nonumber\\\left.+G^{\mx{\Omega}}_J\sum_{m=0}^{\min(K_I,K_J)}\left(\frac{p_{IJ}p_{IJ}}{p_{II}p_{JJ}}\right)^m H_1(m,K_I,K_J,L_I)\right]\label{eq:omegaIJ}
\end{gather}
where $\beta\in\{+,-,z\}$, $\mx{A}_{IJ}=\mx{A}_I+\mx{A}_J$ and the terms related to the overlap of the ECGs are
\begin{gather}
 p_{XY}=\frac{1}{2}\mx{u}_X\mx{A}_{IJ}^{-1}\mx{u}_Y\quad\text{with}\quad X,Y\in\{I,J\}\ .
\end{gather}
The length and velocity dependence is contained within the factors
\begin{gather}
G^\mx{\Omega}_I =\begin{cases}
        \displaystyle\sum_n\frac{q_n(\fett{A}_J\fett{A}_{IJ}^{-1}\fett{v}_I)_n}{m_n(E_{\mr{i}}-E_{\mr{f}})}\quad&\text{if $\mx{\Omega}=\fett{\mu}^\mr{(v)}_\mr{if}$}\\
        \displaystyle\sum_n q_n(\fett{A}_{IJ}^{-1}\fett{v}_I)_n&\text{if $\mx{\Omega}=\fett{\mu}^\mr{(l)}_\mr{if}$}
              \end{cases}\label{eq:GI}\\
G^\mx{\Omega}_J =\begin{cases}
        \displaystyle\sum_n \frac{q_n(\fett{A}_J\fett{A}^{-1}_{IJ}\fett{v}_J-\fett{v}_J)_n}{m_n(E_{\mr{i}}-E_{\mr{f}})}&\text{if $\mx{\Omega}=\fett{\mu}^\mr{(v)}_\mr{if}$}\\
        \displaystyle\sum_n q_n(\fett{A}_{IJ}^{-1}\fett{v}_{J})_n&\text{if $\mx{\Omega}=\fett{\mu}^\mr{(l)}_\mr{if}$}
              \end{cases}\label{eq:GJ}
\end{gather}
while the different contributions from the ``$x$'', ``$y$'' and ``$z$'' components are contained within the factors
\begin{gather}
  C^\beta_1(L_J,L_I,M_{LJ},M_{LI})=\begin{cases}
        \left(\frac{(L_J+M_{LJ}+1)(L_J-M_{LJ}+1)}{(2L_J+1)(2L_J+3)}\right)^{1/2}\delta_{M_{LJ},M_{LI}}\delta_{L_I,L_J+1}&\text{if $\beta=z$}\\
        \left(\frac{(L_J-M_{LJ}+2)(L_J-M_{LJ}+1)}{(2L_J+1)(2L_J+3)}\right)^{1/2}\delta_{M_{LJ},M_{LI}+1}\delta_{L_I,L_J+1}&\text{if $\beta=+$}\\
        -\left(\frac{(L_J+M_{LJ}+2)(L_J+M_{LJ}+1)}{(2L_J+1)(2L_J+3)}\right)^{1/2}\delta_{M_{LI},M_{LJ}+1}\delta_{L_I,L_J+1}&\text{if $\beta=-$}\\
              \end{cases}\\
 C^\beta_2(L_I,L_J,M_L,M_{LJ})=\begin{cases}
        \left(\frac{(L_I+M_{LI}+1)(L_I-M_{LI}+1)}{(2L_I+1)(2L_I+3)}\right)^{1/2}\delta_{M_{LI},M_{LJ}}\delta_{L_J,L_I+1}&\text{if $\beta=z$}\\
        -\left(\frac{(L_I+M_{LI}+2)(L_I+M_{LI}+1)}{(2L_I+1)(2L_I+3)}\right)^{1/2}\delta_{M_{LJ},M_{LI}+1}\delta_{L_J,L_I+1}&\text{if $\beta=+$}\\
        \left(\frac{(L_I-M_{LI}+2)(L_I-M_{LI}+1)}{(2L_I+1)(2L_I+3)}\right)^{1/2}\delta_{M_{LI},M_{LJ}+1}\delta_{L_J,L_I+1}&\text{if $\beta=-$}\\
              \end{cases}
\end{gather}
which include the selection rules for the different transitions. Note that the selection rules emerge naturally from the derivation and are not included in a technical fashion. In order to increase the efficiency of the calculations the following terms are precalculated:
\begin{gather}
 H_1(m,K_1,K_2,L)=\frac{B_{mL}}{(K_1-m)!(K_2-m)!(2m+L)!}\left[F(K_1,L+1)F(K_2,L)\right]^{-1/2}\\
 H_2(m,K_1,K_2,L)=\frac{B_{(m-1)L}}{(K_1-m+1)!(K_2-m)!(2m+L-2)!}\left[F(K_1,L)F(K_2,L-1)\right]^{-1/2}\ .
\end{gather}
The remaining terms
\begin{gather}
 q_X=\frac{1}{2}\mx{u}_X\trs\mx{A}_X^{-1}\mx{u}_X\quad\text{with}\quad X\in\{I,J\}\\
F(K,L)=\sum_{m=0}^K\frac{4^mB_{mL}}{4^K(K-m)!(K-m)!(L+2m)!}
\end{gather}
originate from the quasi-normalization.

Finally, we perform the back transformations of Eqs.\ (\ref{eq:omegap}) and (\ref{eq:omegam}) in order to obtain the original $(x,y,z)$ Cartesian components
\begin{gather}
  C^\beta_1(L_J,L_I,M_{LJ},M_{LI})=\begin{cases}
        \frac{1}{2}[C^-_1(L_J,L_I,M_{LJ},M_{LI})+C^+_1(L_J,L_I,M_{LJ},M_{LI})]&\text{if $\beta=x$}\\
        \frac{i}{2}[C^-_1(L_J,L_I,M_{LJ},M_{LI})-C^+_1(L_J,L_I,M_{LJ},M_{LI})]&\text{if $\beta=y$}\\
              \end{cases}\ ,\\
 C^\beta_2(L_I,L_J,M_L,M_{LJ})=\begin{cases}
        \frac{1}{2}[C^-_2(L_I,L_J,M_L,M_{LJ})+C^+_2(L_I,L_J,M_L,M_{LJ})]&\text{if $\beta=x$}\\
        \frac{i}{2}[C^-_2(L_I,L_J,M_L,M_{LJ})-C^+_2(L_I,L_J,M_L,M_{LJ})]&\text{if $\beta=y$}\\
              \end{cases}\ .
\end{gather}

This step concludes the derivation for the integral expressions for the electric transition dipole moment. The squared length is then calculated according to Eq.\ (\ref{eq:mu}) as the sum of the squared lengths of the allowed transitions among the degenerate substates of the initial and final state.

\subsection{Elimination of the Translational Contamination}

We aim at calculating an internal molecular property free from contributions of the center of mass Cartesian coordinate. The traditional approach is to perform a linear transformation $\mx{U}_x$ of the LFCC
\begin{align}
  \mx{x}_\mathrm{TICM} =
  \left[%
    \begin{array}{@{}c@{}}
      \mx{x} \\
      \mx{X}_\mr{CM} \\
    \end{array}
  \right]
  =
  (\mx{U}_x \otimes \mx{1}_3) \mx{r}
  \label{eq:defticm}
\end{align}
which separates the coordinates of the center of mass, $\mx{X}_\mr{CM}$, from a set of translationally invariant Cartesian coordinates $\fett{x}^{\mr{T}}=(\fett{x}_1^{\mr{T}},\ldots,\fett{x}_n^{\mr{T}})$ (TICC) \cite{Sutcliffe2003}. The Hamiltonian and any property operator are transformed accordingly. The separated center of mass coordinate can then be eliminated and a new formalism is obtained which only involves TICC.

As an alternative, we showed in a recent paper\cite{Simmen2013} that it is possible to avoid any coordinate transformation and work with the original LFCC. The translational contamination is identified in the integral expressions and eliminated by subtraction. This practical approach is based on a parametrization of the basis functions considering the transformation behaviour of $\mx{A}$ and $\mx{u}$ between LFCC and TICC:
\begin{gather}
 \mx{A}=\mx{U}_x^{\mr{T}}\mx{A}^{(x)}\mx{U}_x\quad\text{and}\quad\mx{u}=\mx{U}_x^{\mr{T}}\mx{u}^{(x)}\ .\label{eq:aux}
\end{gather}
The transformed $\mx{A}^{(x)}$ and $\mx{u}^{(x)}$ feature a special block structure:
\begin{align}
  \mx{A}^{(x)}
  =
  \left[%
    \begin{array}{@{}cc@{}}
      \mathpzc{A}^{(x)} &   0 \\
                      0 & c_A \\
    \end{array}
  \right]
  \quad
  \text{and}
  \quad
  \mx{u}^{(x)}
  =
  \left[%
    \begin{array}{@{}c@{}}
      \mathpzc{u}^{(x)} \\
                    c_u \\
    \end{array}
  \right] \ .
  \label{eq:aublock}
\end{align}
The factors $c_A$ and $c_u$ can be assigned to the Cartesian coordinates of the center of mass. The basis functions are   translationally invariant if both $c_A$ and $c_u$ are zero\cite{Matyus2012}. For $c_u$, this does not cause any problems. Yet, $c_A=0$ is in direct contradiction with the requirement that $\mx{A}$ is positive-definite which is important to ensure a non-vanishing (positive-definite) norm for the wave function. Hence, we have to select $c_A>0$ for any practical implementation and subtract the contributions of $c_A$ from the individual terms in the integral expression.

The contributions from $c_A$ to Eq.\ (\ref{eq:omegaIJ}) are identified by substituting $\mx{A}$ and $\mx{u}$ with their corresponding expressions in terms of $\mx{A}^{(x)}$ and $\mx{u}^{(x)}$ according to Eq.\ (\ref{eq:aux}). Most of the contributions of $c_A$ cancel from the integral expressions due to the quasi-normalization of the basis functions. The remaining contributions of $c_A$ are then eliminated from the integral expressions by subtraction if necessary. There are two terms through which such a contamination might be introduced in the dipole integrals: $G^\Omega_I$, Eq.\ (\ref{eq:GI}), and $G^\Omega_J$, Eq.\ (\ref{eq:GJ}). Performing the substitutions, we find that the results of Eq.\ (\ref{eq:omegaIJ}) are free from any translational contamination if $c_u$ is zero. This allows us to evaluate and implement the transition dipole integrals in the LFCC formalism.

The advantages of performing calculations in LFCC are three-fold. First, one does not have to choose a set of TICC, which introduces some ambiguity. Furthermore the physical picture is more intuitive, and most importantly, the many-particle integrals are evaluated more easily\cite{Kozlowski1993}. An advantage of a TICC pre-BO framework is that the dimension of $\mx{A}$ is $n\times n$ rather than $(n+1)\times (n+1)$ so some of the involved matrix operations become slightly computationally less expensive. TICCs are also more suited to represent correlations paths among certain particles\cite{Matyus2012}.

Since the parameters $c_A$ and $c_u$ are independent for the specific choice of the transformation $\mx{U}_x$, LFCC-TICC hybrid methods can be imagined where specific correlations paths can be included into the parametrization. This idea might be investigated in later work.

The numerical implementation of the dipole integrals has been validated for HT$^+$ by comparison to the results of 
Tian and co-workers\cite{Tian2012} as references.

\section{Numerical Results}
\label{sec:results}

\begin{figure}[H]
 \begin{center}
 \includegraphics[width=\textwidth]{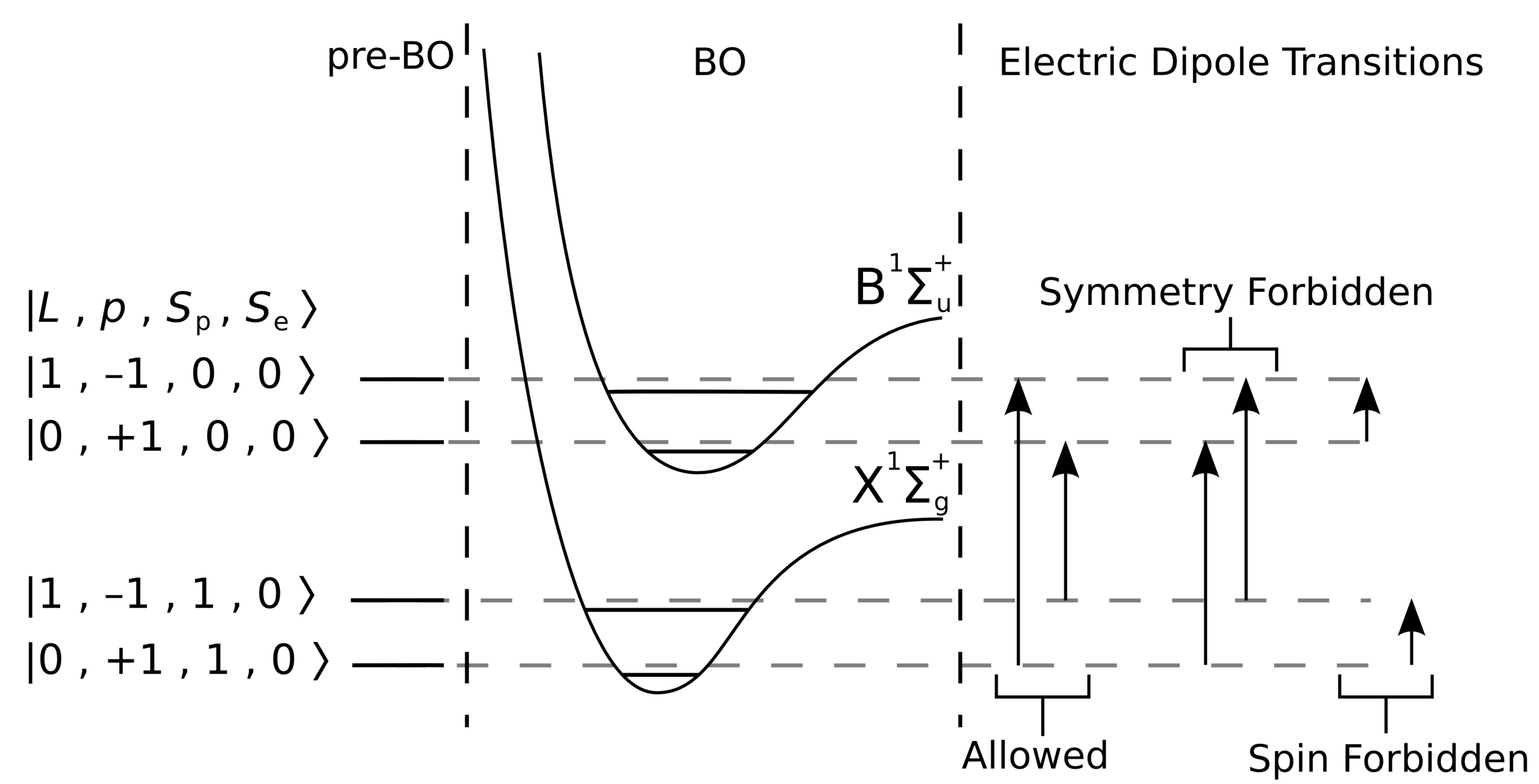}
 \end{center}
 \caption{\label{fig:H2Trans}\small Sketch of the spectrum of H$_2=\lbrace\mr{p}^+,\mr{p}^+,\mr{e}^-,\mr{e}^-\rbrace$ for the lowest two rovibronic states involved in this work.  All states are illustrated in terms of the BO and the pre-BO framework. The individual electronic states are designated by their electronic state labels $X\ ^1\Sigma^+_\mr{g}$ and $B\ ^1\Sigma^+_\mr{u}$ and the relevant quantum numbers ($L$: spatial angular momentum state; $p$: parity $(-1)^L$; $S_\mr{e}$: electronic spin state; $S_\mr{p}$: proton spin state). Furthermore, all potential transitions are listed. We denote whether the transitions are allowed or forbidden with respect to electric dipole transitions. Note that the pre-BO energy levels are generally higher than the corresponding BO energy levels.}
\end{figure}

In this section, we present numerical results for the electric transition dipole moment calculated for the H$_2=\lbrace\mr{p}^+,\mr{p}^+,\mr{e}^-,\mr{e}^-\rbrace$ molecule. Pure rotational dipole transitions of H$_2$ with $\Delta L=\pm 1$ are not possible because of the alternating ortho ($S_\mathrm{p}=1$) and para ($S_\mathrm{p}=0$) states in the ground electronic state\cite{Woolley1976}. There can be, however, non-vanishing transition moments between rovibronic levels of different electronic states with the same electron spin state ($S_\mathrm{e}$). Thus, we have considered rovibronic transitions between energy levels assignable to the two lowest-lying singlet electronic states in the BO theory, $X\ ^1\Sigma^+_\mr{g}$ and $B\ ^1\Sigma^+_\mr{u}$ (Figure \ref{fig:H2Trans}). For these transitions the electronic dipole transition function has already been calculated \cite{Wolniewicz2003}, but we are not aware of any calculation of the dipole transition moments using this dipole transition function in a non-adiabatic framework. In the present work, we do not rely on any dipole moment function, but evaluate the electric transition dipole moments by directly evaluating the transition dipole integrals with the pre-BO wave functions. The wave functions used in this work have been successfully applied for the calculation of resonances in a recent work\cite{Matyus2013}. The employed parameter sets corresponding to 2250 basis functions for each state are provided in the supporting information. The mass of the proton $m_\mr{p}$ was chosen in terms of the electron mass $m_\mr{e}$ as $m_\mr{p}/m_\mr{e}=1836.15267247$\cite{Mohr2008}.

\begin{table}[H]
 \caption{\label{tab:H2energies}\small Energies $E$ of the H$_2=\lbrace\mr{p}^+,\mr{p}^+,\mr{e}^-,\mr{e}^-\rbrace$ parameter sets used in this work together with references from the literature. The size of the parameter sets was 2250 in all cases. All results are in Hartree atomic units [\Eh]. Only vibrational ground states are considered.}
 \begin{center}
 \renewcommand{\baselinestretch}{1.0}
 \renewcommand{\arraystretch}{1}
\begin{tabular}{cccccc}
\hline
\hline
($L$,$p$,$S_\mr{p}$,$S_\mr{e}$)$^a$&$E$ [\Eh]$^b$&$\eta$ [$10^{-9}$]$^c$&$E_{\mathrm{Ref}}-E$ [$\mu$\Eh]&Ref.&Assignment$^d$\\
\hline
(0,+1,0,0)&$-$1.164025029&1.451&$-$0.0014&\cite{Pachucki2009a}&$X\ ^1\Sigma^+_\mr{g}$\\
(1,$-$1,1,0)&$-$1.163485171&2.217&$-$0.0014&\cite{Pachucki2009a}&$X\ ^1\Sigma^+_\mr{g}$\\
\hline
(0,+1,1,0)&$-$0.753027184&7.714&1.3813&\cite{Wolniewicz2006}&$B\ ^1\Sigma^+_\mr{u}$\\
(1,$-$1,0,0)&$-$0.752850232&1.515&1.4435&\cite{Wolniewicz2006}&$B\ ^1\Sigma^+_\mr{u}$\\
\hline
\hline
  \end{tabular}
 \end{center}
\begin{flushleft}
\small
$a$: $L$: Spatial angular momentum quantum number; $p$: parity ($p=(-1)^L$); $S_\mr{p}$ and $S_\mr{e}$: total spin quantum numbers for the protons and the
 electrons, respectively.\\
$b$: Energy obtained from the parameter set used in a recent work \cite{Matyus2013}. The parameter sets are available in the supporting information.\\
$c$: The virial ratio: $\eta=|1+\bra{\Psi}\hat{V}\ket{\Psi}/(2\bra{\Psi}\hat{T}\ket{\Psi})|$ where $\bra{\Psi}\hat{T}\ket{\Psi}$ and $\bra{\Psi}\hat{V}\ket{\Psi}$ are the kinetic and the potential energies respectively. In the case of the exact wave function we find $\eta=0$.\\
$d$: Born--Oppenheimer electronic state label. Each energy level given here can be assigned to the lowest-energy vibrational level of the electronic state.
\end{flushleft}
\end{table}

Table \ref{tab:H2energies} lists the energies corresponding to the basis functions which we use for the rovibronic states considered. We use reference values from the literature to assess the quality of our parameter sets. Our energies are either comparable to the best available reference values ($X\ ^1\Sigma^+_\mr{g}$) or more accurate ($B\ ^1\Sigma^+_\mr{u}$) and therefore suited for the calculation of electric transition dipole moments.

\begin{table}[H]
 \caption{\label{tab:H2transitions}\small Electric transition dipole moments obtained in our work for H$_2=\lbrace\mr{p}^+,\mr{p}^+,\mr{e}^-,\mr{e}^-\rbrace$. The initial and final states are characterized by their assigned state label and the associated quantum numbers. All results are in atomic units. Only vibrational ground states are considered.}
 \begin{center}
 \renewcommand{\baselinestretch}{1.0}
 \renewcommand{\arraystretch}{1}
\begin{tabular}{ccccccc}
\hline
\hline
\multicolumn{2}{c}{Initial State}&\multicolumn{2}{c}{Final State}&&\multicolumn{2}{c}{Transition Dipole Moments}\\
\hline
Assignment$^a$&($L$,$p$,$S_\mr{p}$,$S_\mr{e}$)$^b$&Assignment$^a$&($L$,$p$,$S_\mr{p}$,$S_\mr{e}$)$^b$&E$_\mr{i}$-E$_\mr{f}$ [\Eh]&$|\fett{\mu}_\mr{if}^\mr{(v)}|$&$|\fett{\mu}_\mr{if}^\mr{(l)}|$\\
\hline
$X\ ^1\Sigma^+_\mr{g}$&(0,+1,0,0)&$B\ ^1\Sigma^+_\mr{u}$&(1,--1,0,0)&--0.411174797&0.078415&0.078414\\
\hline
$B\ ^1\Sigma^+_\mr{u}$&(0,+1,1,0)&$X\ ^1\Sigma^+_\mr{g}$&(1,--1,1,0)&--0.410457987&0.079802&0.079801\\
\hline
\hline
  \end{tabular}
 \end{center}
\begin{flushleft}
\small
$a$: Born--Oppenheimer electronic state label. Each energy level given here can be assigned to the lowest-energy vibrational level of the electronic state.\\
$b$: $L$: Spatial angular momentum quantum number; $p$: parity ($p=(-1)^L$); $S_\mr{p}$ and $S_\mr{e}$: total spin quantum numbers for the protons and the
 electrons, respectively.\\
\end{flushleft}
\end{table}

The calculated electric transition dipole moments are listed in Table \ref{tab:H2transitions}. The transition dipole moments are presented with a precision that shows the first differing digit. We recognize that the values for $|\fett{\mu}_\mr{if}^\mr{(l)}|$ and $|\fett{\mu}_\mr{if}^\mr{(v)}|$ have converged.

\section{Conclusions}
We presented the expressions for the integrals of the electric transition dipole moments and its squared length. We exploited the simple form of the electric transition dipole operators in laboratory-fixed Cartesian coordinates. Integral expressions were derived for the components of the transition dipole in the length and the velocity representation. These two representations only yield identical results for exact wave functions.

We have then calculated the electric transition dipole moments for the H$_2$ molecule for transitions between the lowest two rovibronic levels of the ortho- and para-H$_2$. Some sensitivity to the approximation of the wave functions was observed. Yet this was negligible and we obtained converged results for the electric transition dipole moments.

Furthermore, we illustrated the strength of our scheme for the elimination of the translational contribution to any internal molecular property. This scheme was presented in our previous work\cite{Simmen2013} and allows us to perform pre-Born--Oppenheimer calculations in laboratory-fixed Cartesian coordinates. Previous work relied on a linear combination of these LFCCs to yield a set of translationally invariant Cartesian coordinates, which separates the center of mass coordinate in order to eliminate any contribution from the overall motion of the system.

\section{Acknowledgments}
This work has been supported by the Swiss National Science Foundation. EM thanks the Hungarian Scientific Research Fund (OTKA, NK83583) for financial support.

\small

\begin{thebibliography}{10}

\bibitem{Bubin2012}
{\sc S.~Bubin}, {\sc M.~Pavanello}, {\sc W.-C. Tung}, {\sc K.~L. Sharkey}, and
  {\sc L.~Adamowicz},
\newblock {\em Chem. Rev.} {\bf 113}, 36 (2013).

\bibitem{Mitroy2013}
{\sc J.~Mitroy}, {\sc S.~Bubin}, {\sc W.~Horiuchi}, {\sc Y.~Suzuki}, {\sc
  L.~Adamowicz}, {\sc W.~Cencek}, {\sc K.~Szalewicz}, {\sc J.~Komasa}, {\sc
  D.~Blume}, and {\sc K.~Varga},
\newblock {\em Rev. Mod. Phys.} {\bf 85}, 693 (2013).

\bibitem{Varga1998}
{\sc K.~Varga}, {\sc Y.~Suzuki}, and {\sc J.~Usukura},
\newblock {\em Few-Body Syst.} {\bf 24}, 81 (1998).

\bibitem{Suzuki1998}
{\sc Y.~Suzuki}, {\sc J.~Usukura}, and {\sc K.~Varga},
\newblock {\em J. Phys. B} {\bf 31}, 31 (1998).

\bibitem{Suzuki1998a}
{\sc Y.~Suzuki} and {\sc K.~Varga},
\newblock {\em {Stochastic Variational Approach to Quantum-Mechanical Few-Body
  Problems}}, volume~54 of {\em Lect. Notes Phys.},
\newblock Springer, Berlin, Heidelberg, 1998.

\bibitem{Matyus2012}
{\sc E.~M\'{a}tyus} and {\sc M.~Reiher},
\newblock {\em J. Chem. Phys.} {\bf 137}, 024104 (2012).

\bibitem{Simmen2013}
{\sc B.~Simmen}, {\sc E.~M\'{a}tyus}, and {\sc M.~Reiher},
\newblock {\em Mol. Phys.} {\bf 111}, 2086 (2013).

\bibitem{Matyus2013}
{\sc E.~M\'{a}tyus},
\newblock {\em J. Phys. Chem. A} {\bf 117}, 7195 (2013).

\bibitem{Jeziorski1979}
{\sc B.~Jeziorski} and {\sc K.~Szalewicz},
\newblock {\em Phys. Rev. A} , 2360 (1979).

\bibitem{Cencek1993}
{\sc W.~Cencek} and {\sc J.~Rychlewski},
\newblock {\em J. Chem. Phys.} {\bf 98}, 1252 (1993).

\bibitem{Rychlewski2004}
{\sc J.~Rychlewski}, editor,
\newblock {\em {Explicitly Correlated Wave Functions in Chemistry and
  Physics}},
\newblock Kluwer Academic Publishers, 2003.

\bibitem{Boys1960}
{\sc S.~F. Boys},
\newblock {\em Proc. R. Soc. London Ser. A} {\bf 258}, 402 (1960).

\bibitem{Singer1960}
{\sc K.~Singer},
\newblock {\em Proc. R. Soc. London Ser. A} {\bf 258}, 412 (1960).

\bibitem{Tian2012}
{\sc Q.-L. Tian}, {\sc L.-Y. Tang}, {\sc Z.-X. Zhong}, {\sc Z.-C. Yan}, and
  {\sc T.-Y. Shi},
\newblock {\em J. Chem. Phys.} {\bf 137}, 024311 (2012).

\bibitem{Bekbaev2011}
{\sc A.~K. Bekbaev}, {\sc V.~I. Korobov}, and {\sc M.~Dineykhan},
\newblock {\em Phys. Rev. A} {\bf 83}, 044501 (2011).

\bibitem{Ion}
{\sc E.~R. Cohen}, {\sc T.~Cvitas}, {\sc J.~G. Frey}, {\sc B.~Holmstr\"{o}m},
  {\sc K.~Kochitsu}, {\sc R.~Marquart}, {\sc I.~Mills}, {\sc F.~Pavese}, {\sc
  M.~Quack}, {\sc J.~Stohner}, {\sc H.~L. Strauss}, {\sc M.~Tamaki}, and {\sc
  A.~J. Thor},
\newblock {\em {Quantities, Units and Symbols in Physical Chemistry}},
\newblock 3rd edition, 2007.

\bibitem{Anderson1999}
{\sc E.~Anderson}, {\sc Z.~Bai}, {\sc C.~Bischof}, {\sc S.~Blackford}, {\sc
  J.~Demmel}, {\sc J.~Dongarra}, {\sc J.~{Du Croz}}, {\sc A.~Greenbaum}, {\sc
  S.~Hammarling}, {\sc A.~McKenney}, and {\sc D.~Sorensen},
\newblock {\em {LAPACK Users' Guide}},
\newblock Society for Industrial and Applied Mathematic, Philadelphia, PA,
  third edition, 1999.

\bibitem{Sanderson2010}
{\sc C.~Sanderson},
\newblock {Armadillo: An open source C++ linear algebra library for fast
  prototyping and computationally intensive experiments},
\newblock Technical report, NICTA, 2010.

\bibitem{Matyus2011a}
{\sc E.~M\'{a}tyus}, {\sc J.~Hutter}, {\sc U.~M\"{u}ller-Herold}, and {\sc
  M.~Reiher},
\newblock {\em J. Chem. Phys.} {\bf 135}, 204302 (2011).

\bibitem{Matyus2011}
{\sc E.~M\'{a}tyus}, {\sc J.~Hutter}, {\sc U.~M\"{u}ller-Herold}, and {\sc
  M.~Reiher},
\newblock {\em Phys. Rev. A} {\bf 83}, 052512 (2011).

\bibitem{Rodriguez2013}
{\sc C.~G. Rodr\'{\i}guez}, {\sc A.~S. Urbina}, {\sc F.~J. Torres}, {\sc
  D.~Cazar}, and {\sc E.~V. Lude\~{n}a},
\newblock {\em Comp. Theor. Chem.} {\bf 1018}, 26 (2013).

\bibitem{Muller-Herold2006}
{\sc U.~M\"{u}ller-Herold},
\newblock {\em J. Chem. Phys.} {\bf 124}, 14105 (2006).

\bibitem{Muller-Herold2008}
{\sc U.~M\"{u}ller-Herold},
\newblock {\em Eur. Phys. J. D} {\bf 49}, 311 (2008).

\bibitem{Ludena2012}
{\sc E.~V. Lude\~{n}a}, {\sc L.~Echevarr\'{\i}a}, {\sc X.~Lopez}, and {\sc
  J.~M. Ugalde},
\newblock {\em J. Chem. Phys.} {\bf 136}, 084103 (2012).

\bibitem{Becerra2013}
{\sc M.~Becerra}, {\sc V.~Posligua}, and {\sc E.~V. Lude\~{n}a},
\newblock {\em Int. J. Quantum Chem.} {\bf 113}, 1584 (2013).

\bibitem{Goli2012}
{\sc M.~Goli} and {\sc S.~Shahbazian},
\newblock {\em Theor. Chem. Acc.} {\bf 131}, 1208 (2012).

\bibitem{Goli2013}
{\sc M.~Goli} and {\sc S.~Shahbazian},
\newblock {\em Theor. Chem. Acc.} {\bf 132}, 1410 (2013).

\bibitem{Aguirre2013}
{\sc N.~F. Aguirre}, {\sc P.~Villarreal}, {\sc G.~Delgado-Barrio}, {\sc
  E.~Posada}, {\sc A.~Reyes}, {\sc M.~Biczysko}, {\sc A.~O. Mitrushchenkov},
  and {\sc M.~P. de~Lara-Castells},
\newblock {\em J. Chem. Phys} {\bf 138}, 184113 (2013).

\bibitem{King2013}
{\sc A.~W. King}, {\sc F.~Longford}, and {\sc H.~Cox},
\newblock {\em J. Chem. Phys} {\bf 139}, 224306 (2013).

\bibitem{Perez-Torres2013}
{\sc J.~F. P\'{e}rez-Torres},
\newblock {\em Phys. Rev. A} {\bf 87}, 062512 (2013).

\bibitem{Chakraborty2008}
{\sc A.~Chakraborty}, {\sc M.~V. Pak}, and {\sc S.~Hammes-Schiffer},
\newblock {\em J. Chem. Phys.} {\bf 129}, 014101 (2008).

\bibitem{Sutcliffe2005a}
{\sc B.~T. Sutcliffe} and {\sc R.~G. Woolley},
\newblock {\em Chem. Phys. Lett.} {\bf 408}, 445 (2005).

\bibitem{Sutcliffe2005b}
{\sc B.~T. Sutcliffe} and {\sc R.~G. Woolley},
\newblock {\em Phys. Chem. Chem. Phys.} {\bf 7}, 3664 (2005).

\bibitem{Woolley1976}
{\sc R.~G. Woolley},
\newblock {\em Adv. Phys.} {\bf 25}, 27 (1976).

\bibitem{Woolley1998}
{\sc R.~G. Woolley},
\newblock {\em J. Math. Chem.} {\bf 23}, 3 (1998).

\bibitem{Woolley1991}
{\sc R.~G. Woolley},
\newblock {\em Comp. Theor. Chem.} {\bf 230}, 17 (1991).

\bibitem{Cafiero2003}
{\sc M.~Cafiero}, {\sc S.~Bubin}, and {\sc L.~Adamowicz},
\newblock {\em Phys. Chem. Chem. Phys.} {\bf 5}, 1491 (2003).

\bibitem{Hilborn1982}
{\sc R.~C. Hilborn},
\newblock {\em Am. J. Phys.} {\bf 50}, 982 (1982).

\bibitem{Chandrasekhar1945}
{\sc S.~Chandrasekhar},
\newblock {\em Astrophys. J.} {\bf 102}, 223 (1945).

\bibitem{Anderson1974}
{\sc M.~T. Anderson} and {\sc F.~Weinhold},
\newblock {\em Phys. Rev. A} {\bf 10}, 1457 (1974).

\bibitem{Sutcliffe2003}
{\sc B.~T. Sutcliffe},
\newblock {Chapter 31 Coordinate Systems and Transformations},
\newblock in {\em Handbook of Molecular Physics and Quantum Chemistry}, edited
  by {\sc S.~Wilson}, p. 485, John Wiley \& Sons, Ltd, Chichester, 2003.

\bibitem{Kozlowski1993}
{\sc P.~M. Kozlowski} and {\sc L.~Adamowicz},
\newblock {\em Chem. Rev.} {\bf 93}, 2007 (1993).

\bibitem{Wolniewicz2003}
{\sc L.~Wolniewicz} and {\sc G.~Staszewska},
\newblock {\em J. Mol. Spectrosc.} {\bf 217}, 181 (2003).

\bibitem{Mohr2008}
{\sc P.~J. Mohr}, {\sc B.~N. Taylor}, and {\sc D.~B. Newell},
\newblock {\em Rev. Mod. Phys.} {\bf 80}, 633 (2008).

\bibitem{Pachucki2009a}
{\sc K.~Pachucki} and {\sc J.~Komasa},
\newblock {\em J. Chem. Phys.} {\bf 130}, 164113 (2009).

\bibitem{Wolniewicz2006}
{\sc L.~Wolniewicz}, {\sc T.~Orlikowski}, and {\sc G.~Staszewska},
\newblock {\em J. Mol. Spectrosc.} {\bf 238}, 118 (2006).

\end{thebibliography}




\section*{Supplementary material}

# Electric Transition Dipole Moment in pre-Born-Oppenheimer Molecular Structure Theory

Benjamin Simmen[a], Edit Mátyus[b1], Markus Reiher[a2]

[a]ETH Zürich, Laboratorium für Physikalische Chemie,
Vladimir-Prelog-Weg 2, 8093 Zürich, Switzerland

[b]Eötvös University, Institute of Chemistry, P.O. Box 32, H-1518, Budapest 112, Hungary
Present Address: Department of Chemistry, University of Cambridge, Lensfield Road, Cambridge, CB2 1EW, United Kingdom

**Supporting Information**

This supporting information contains the derivation of the integral expressions for the transition dipole operator for both the length and velocity operator. The basis functions considered are explicitly correlated Gaussian functions in the global vector representation. The radial integral is performed over a set of Laboratory-Fixed Cartesian coordinates $r$. The angular integral is evaluated over a set of coordinates with components transformed in accordance with the spin ladder operators where $\varepsilon_x$, $\varepsilon_y$ and $\varepsilon_z$ are unit direction vectors:

---

[1]corresponding author; e-mail: matyus@chem.elte.hu
[2]corresponding author; e-mail: markus.reiher@phys.chem.ethz.ch

# Contents

# 1 Introduction

In this section, we start with the introduction of the coordinate transformation used for the evaluation of the angular integrals. Then, we present the three step integral evaluation procedure used in this derivation.

## 1.1 Transformed Coordinates

For the evaluation of the angular integrals we use the following transformed coordinates:

$$\varepsilon_- = \varepsilon_x + i\varepsilon_y \tag{S1}$$

$$\varepsilon_+ = \varepsilon_x - i\varepsilon_y \tag{S2}$$

while $\varepsilon_z$ is the same in both coordinate systems. This change of the coordinate system yields simple expressions for the angular integrals. The operators of the momentum and position components are transformed accordingly

$$\mathbf{\Omega}_- = \mathbf{\Omega}_x + i\mathbf{\Omega}_y \tag{S3}$$

$$\mathbf{\Omega}_+ = \mathbf{\Omega}_x - i\mathbf{\Omega}_y \tag{S4}$$

with $\mathbf{\Omega} \in \{\boldsymbol{\mu}_{\mathrm{if}}^{(\mathrm{v})}, \boldsymbol{\mu}_{\mathrm{if}}^{(\mathrm{l})}\}$ where

$$\boldsymbol{\mu}_{\mathrm{if}}^{(\mathrm{l})} = \sum_{j=1}^{n+1} \langle \Psi_{\mathrm{i}} | q_j \boldsymbol{r}_j | \Psi_{\mathrm{f}} \rangle \tag{S5}$$

is the length representation of the transition dipole operator and

$$\boldsymbol{\mu}_{\mathrm{if}}^{(\mathrm{v})} = -\frac{1}{(E_{\mathrm{i}} - E_{\mathrm{f}})} \sum_{j=1}^{n+1} \left\langle \Psi_{\mathrm{i}} \left| \frac{q_j}{m_j} \boldsymbol{\nabla}_{\boldsymbol{r}_j} \right| \Psi_{\mathrm{f}} \right\rangle . \tag{S6}$$

is the velocity representation. The components $\mathbf{\Omega}_x$ and $\mathbf{\Omega}_y$ can be expressed as

$$\mathbf{\Omega}_x = \frac{1}{2}(\mathbf{\Omega}_- + \mathbf{\Omega}_+) \tag{S7}$$

$$\mathbf{\Omega}_y = -\frac{i}{2}(\mathbf{\Omega}_- - \mathbf{\Omega}_+) . \tag{S8}$$

The components of the two coordinate systems are collected in the subscripts

$$\alpha \in \{+, -, z\} \tag{S9}$$

$$\beta \in \{x, y, z\} \tag{S10}$$

## 1.2 Three-Step Integral Evaluation Procedure

The integrals expression for some operator $\hat{O}$ and explicitly correlated Gaussian functions with the global vector representation (Eq. (TK) in the paper) can be evaluated following a three step procedure. During the derivation, we rely on the generating functions $g(\boldsymbol{r}, \boldsymbol{A}, a\boldsymbol{u} \otimes \boldsymbol{\varepsilon})$ of the basis functions (Eq. (TK) in the paper). The three steps are then:

1. Evaluate the integrals of $\hat{O}$ with the generating functions:

$$I_1 = \langle g(\boldsymbol{r}, \boldsymbol{A}_I, a_I \boldsymbol{u}_I \otimes \boldsymbol{\varepsilon}_I) | \hat{O} | g(\boldsymbol{r}, \boldsymbol{A}_J, a_J \boldsymbol{u}_J \otimes \boldsymbol{\varepsilon}_J) \rangle \ . \tag{S11}$$

2. Evaluate the derivatives at $a_I = a_J = 0$:

$$I_2 = \frac{\partial^{2K_I+L_I}}{\partial a_I^{2K_I+L_I}} \frac{\partial^{2K_J+L_J}}{\partial a_J^{2K_J+L_J}} I_1(a_I, a_J) \bigg|_{a_I=a_J=0} \ . \tag{S12}$$

3. Evaluate the angular integrals:

$$I_3 = \frac{1}{B_{K_I L_I} B_{K_J L_J}} \int d\hat{\boldsymbol{\varepsilon}}_I \int d\hat{\boldsymbol{\varepsilon}}_J Y_{M_{L_I}}^{L_I *}(\hat{\boldsymbol{\varepsilon}}_I) Y_{M_{L_J}}^{L_J}(\hat{\boldsymbol{\varepsilon}}_J) I_2(\boldsymbol{\varepsilon}_I, \boldsymbol{\varepsilon}_J) \ . \tag{S13}$$

# 2 Radial Integrals and Derivative

We start by deriving the radial integral expressions for the one-particle momentum and position operators. The resulting expressions are then used to formulate one general radial integral expression for the transition dipole operator in both the length and velocity representation.

## 2.1 Momentum Components Integral Expressions

$$\left\langle \exp\left(-\frac{1}{2}\boldsymbol{r}^{\mathrm{T}}(\boldsymbol{A}_I \otimes \mathbf{1}_3)\boldsymbol{r} + \boldsymbol{v}_I^{\mathrm{T}}\boldsymbol{r}\right) \middle| \frac{\partial}{\partial (\boldsymbol{r}_n)_\beta} \middle| \exp\left(-\frac{1}{2}\boldsymbol{r}^{\mathrm{T}}(\boldsymbol{A}_J \otimes \mathbf{1}_3)\boldsymbol{r} + \boldsymbol{v}_J^{\mathrm{T}}\boldsymbol{r}\right) \right\rangle$$

$$= \left\langle \exp\left(-\frac{1}{2}\boldsymbol{r}^{\mathrm{T}}(\boldsymbol{A}_I \otimes \mathbf{1}_3)\boldsymbol{r} + \boldsymbol{v}_I^{\mathrm{T}}\boldsymbol{r}\right) \middle| \exp\left(-\frac{1}{2}\boldsymbol{r}^{\mathrm{T}}(\boldsymbol{A}_J \otimes \mathbf{1}_3)\boldsymbol{r} + \boldsymbol{v}_J^{\mathrm{T}}\boldsymbol{r}\right) \times \frac{\partial}{\partial (\boldsymbol{r}_n)_\beta}\left(-\frac{1}{2}\boldsymbol{r}^{\mathrm{T}}(\boldsymbol{A}_J \otimes \mathbf{1}_3)\boldsymbol{r} + \boldsymbol{v}_J^{\mathrm{T}}\boldsymbol{r}\right) \right\rangle$$

$$= \left\langle \exp\left(-\frac{1}{2}\boldsymbol{r}^{\mathrm{T}}(\boldsymbol{A}_I \otimes \mathbf{1}_3)\boldsymbol{r} + \boldsymbol{v}_I^{\mathrm{T}}\boldsymbol{r}\right) \middle| \exp\left(-\frac{1}{2}\boldsymbol{r}^{\mathrm{T}}(\boldsymbol{A}_J \otimes \mathbf{1}_3)\boldsymbol{r} + \boldsymbol{v}_J^{\mathrm{T}}\boldsymbol{r}\right)\left(-\sum_{k=1}^{N}(\boldsymbol{r}_k)_\beta (\boldsymbol{A}_J)_{nk} + (\boldsymbol{v}_J)_{n\beta}\right) \right\rangle$$

$$= \int_{\mathbb{R}^{3N}} d\boldsymbol{r} \exp\left(-\frac{1}{2}\boldsymbol{r}^{\mathrm{T}}(\boldsymbol{A}_{IJ} \otimes \mathbf{1}_3)\boldsymbol{r} + \boldsymbol{v}_{IJ}^{\mathrm{T}}\boldsymbol{r}\right)\left(-\sum_{k=1}^{N}(\boldsymbol{r}_k)_\beta (\boldsymbol{A}_J)_{nk} + (\boldsymbol{v}_J)_{n\beta}\right)$$

$$= \sum_{k=1}^{N}(-(\boldsymbol{A}_J)_{nk}) \int_{\mathbb{R}^{3N}} d\boldsymbol{r} \exp\left(-\frac{1}{2}\boldsymbol{r}^{\mathrm{T}}(\boldsymbol{A}_{IJ} \otimes \mathbf{1}_3)\boldsymbol{r} + \boldsymbol{v}_{IJ}^{\mathrm{T}}\boldsymbol{r}\right)(\boldsymbol{r}_k)_\beta$$

$$+ (\boldsymbol{v}_J)_{n\beta} \int_{\mathbb{R}^{3N}} d\boldsymbol{r} \exp\left(-\frac{1}{2}\boldsymbol{r}^{\mathrm{T}}(\boldsymbol{A}_{IJ} \otimes \mathbf{1}_3)\boldsymbol{r} + \boldsymbol{v}_{IJ}^{\mathrm{T}}\boldsymbol{r}\right)$$

$$= \sum_{k=1}^{N}(-(\boldsymbol{A}_J)_{nk}) D \exp\left(\frac{1}{2}\boldsymbol{v}_{IJ}^{\mathrm{T}}(\boldsymbol{A}_{IJ}^{-1} \otimes \mathbf{1}_3)\boldsymbol{v}_{IJ}\right) \sum_{j=1}^{N}(\boldsymbol{v}_{IJ})_{j\beta}(\boldsymbol{A}_{IJ}^{-1})_{kj}$$

$$+ (\boldsymbol{v}_J)_{n\beta} D \exp\left(\frac{1}{2}\boldsymbol{v}_{IJ}^{\mathrm{T}}(\boldsymbol{A}_{IJ}^{-1} \otimes \mathbf{1}_3)\boldsymbol{v}_{IJ}\right)$$

$$= D \exp\left(\frac{1}{2}\boldsymbol{v}_{IJ}^{\mathrm{T}}(\boldsymbol{A}_{IJ}^{-1} \otimes \mathbf{1}_3)\boldsymbol{v}_{IJ}\right) \sum_{n=1}^{N}\left(-\sum_{k=1}^{N}\sum_{j=1}^{N}(\boldsymbol{A}_J)_{nk}(\boldsymbol{A}_{IJ}^{-1})_{kj}(\boldsymbol{v}_{IJ})_{j\beta} + (\boldsymbol{v}_J)_{n\beta}\right) \quad \text{(S14)}$$

where we have used

$$D = \left(\frac{(2\pi)^N}{\det(A_{IJ})}\right)^{3/2} \quad \text{(S15)}$$

$$\boldsymbol{A}_{IJ} = \boldsymbol{A}_I + \boldsymbol{A}_J \quad \text{(S16)}$$

$$\boldsymbol{v}_{IJ} = \boldsymbol{v}_I + \boldsymbol{v}_J \quad \text{(S17)}$$

for convenience.

Substitution of

$$\boldsymbol{v}_I = a_I\left(\boldsymbol{u}_I \otimes \boldsymbol{\varepsilon}_I\right)$$

$$\boldsymbol{v}_J = a_J\left(\boldsymbol{u}_J \otimes \boldsymbol{\varepsilon}_J\right)$$

into Eq. (S14) yields

$$D \exp\left(a_I^2 p_{II} + a_J^2 p_{JJ} + a_I a_J (\boldsymbol{\varepsilon}_I \boldsymbol{\varepsilon}_J) p_{IJ}\right)\left(a_I (\boldsymbol{\varepsilon}_I)_\beta W_I(n) + a_J (\boldsymbol{\varepsilon}_J)_\beta W_J(n)\right)$$

$$= D \left(a_I (\boldsymbol{\varepsilon}_I)_\beta W_I(n) + a_J (\boldsymbol{\varepsilon}_J)_\beta W_J(n)\right) \sum_{i=0}^{\infty}\sum_{j=0}^{\infty}\sum_{k=0}^{\infty} a_I^{2i+k} a_J^{2j+k} (\boldsymbol{\varepsilon}_I \boldsymbol{\varepsilon}_J)^k \frac{p_{II}^i p_{JJ}^j p_{IJ}^k}{i! j! k!} \quad \text{(S18)}$$

with

$$W_I(n) = -\sum_{k=1}^{N}\sum_{j=1}^{N}(\boldsymbol{A}_J)_{nk}(\boldsymbol{A}_{IJ}^{-1})_{kj}(\boldsymbol{u}_I)_j \tag{S19}$$

$$W_J(n) = -\sum_{k=1}^{N}\sum_{j=1}^{N}(\boldsymbol{A}_J)_{nk}(\boldsymbol{A}_{IJ}^{-1})_{kj}(\boldsymbol{v}_J)_j + \sum_{n=1}^{N}(\boldsymbol{v}_J)_n \tag{S20}$$

$$p_{II} = \frac{1}{2}\boldsymbol{u}_I^{\mathrm{T}}\boldsymbol{A}_{IJ}^{-1}\boldsymbol{u}_I \tag{S21}$$

$$p_{JJ} = \frac{1}{2}\boldsymbol{u}_J^{\mathrm{T}}\boldsymbol{A}_{IJ}^{-1}\boldsymbol{u}_J \tag{S22}$$

$$p_{IJ} = \boldsymbol{u}_I^{\mathrm{T}}\boldsymbol{A}_{IJ}^{-1}\boldsymbol{u}_J \tag{S23}$$

## 2.2 Position Components Integral Elements

$$\left\langle \exp\left(-\frac{1}{2}\boldsymbol{r}^{\mathrm{T}}(\boldsymbol{A}_I\otimes\boldsymbol{1}_3)\boldsymbol{r}+\boldsymbol{v}_I^{\mathrm{T}}\boldsymbol{r}\right)\middle|(\boldsymbol{r}_n)_\beta\middle|\exp\left(-\frac{1}{2}\boldsymbol{r}^{\mathrm{T}}(\boldsymbol{A}_J\otimes\boldsymbol{1}_3)\boldsymbol{r}+\boldsymbol{v}_J^{\mathrm{T}}\boldsymbol{r}\right)\right\rangle$$
$$= \int_{\mathbb{R}^{3N}} d\boldsymbol{r}\exp\left(-\frac{1}{2}\boldsymbol{r}^{\mathrm{T}}(\boldsymbol{A}_{IJ}\otimes\boldsymbol{1}_3)\boldsymbol{r}+\boldsymbol{v}_{IJ}^{\mathrm{T}}\boldsymbol{r}\right)(\boldsymbol{r}_n)_\beta$$
$$= D\exp(\frac{1}{2}\boldsymbol{v}_{IJ}^{\mathrm{T}}\left(\boldsymbol{A}_{IJ}^{-1}\otimes\boldsymbol{1}_3\right)\boldsymbol{v}_{IJ})\sum_{k=1}^{N}(\boldsymbol{A}_{IJ}^{-1})_{nk}(\boldsymbol{v}_{IJ})_{k\beta} \tag{S24}$$

where we have again used

$$D = \left(\frac{(2\pi)^N}{\det(A_{IJ})}\right)^{3/2} \tag{S25}$$

$$\boldsymbol{A}_{IJ} = \boldsymbol{A}_I + \boldsymbol{A}_J \tag{S26}$$

$$\boldsymbol{v}_{IJ} = \boldsymbol{v}_I + \boldsymbol{v}_J \tag{S27}$$

for convenience.

Substitution of

$$\boldsymbol{v}_I = a_I\left(\boldsymbol{u}_I\otimes\boldsymbol{\varepsilon}_I\right)$$
$$\boldsymbol{v}_J = a_J\left(\boldsymbol{u}_J\otimes\boldsymbol{\varepsilon}_J\right)$$

into Eq. (S24) yields

$$D\exp(a_I^2 p_{II} + a_J^2 p_{JJ} + a_I a_J(\boldsymbol{\varepsilon}_I^{\mathrm{T}}\boldsymbol{\varepsilon}_J))\left(R_I a_I(\boldsymbol{\varepsilon}_I)_\beta + R_J a_J(\boldsymbol{\varepsilon}_J)_\beta\right)$$
$$= D\left(a_I(\boldsymbol{\varepsilon}_I)_\beta R_I(n) + a_J(\boldsymbol{\varepsilon}_J)_\beta R_J(n)\right)\sum_{i=0}^{\infty}\sum_{j=0}^{\infty}\sum_{k=0}^{\infty}a_I^{2i+k}a_J^{2j+k}\left(\boldsymbol{\varepsilon}_I\boldsymbol{\varepsilon}_J\right)^k\frac{p_{II}^i p_{JJ}^j p_{IJ}^k}{i!j!k!} \tag{S28}$$

6

with

$$R_I(n) = \sum_{k=1}^{N} (\boldsymbol{A}_{IJ}^{-1})_{nk} (\boldsymbol{u}_I)_k \tag{S29}$$

$$R_J(n) = \sum_{k=1}^{N} (\boldsymbol{A}_{IJ}^{-1})_{nk} (\boldsymbol{u}_J)_k \tag{S30}$$

## 2.3 General Expressions

We can now form a general expression for $I_1(a_I, a_J)$ in step 1 from Eqs. (S18) and (S28)

$$\begin{aligned} I_1(a_I, a_J) &= \left\langle \exp\left(-\frac{1}{2}\boldsymbol{r}^{\mathrm{T}}(\boldsymbol{A}_I \otimes \boldsymbol{1}_3)\boldsymbol{r} + \boldsymbol{v}_I^{\mathrm{T}}\boldsymbol{r}\right) \middle| \boldsymbol{\Omega}_\beta \middle| \exp\left(-\frac{1}{2}\boldsymbol{r}^{\mathrm{T}}(\boldsymbol{A}_J \otimes \boldsymbol{1}_3)\boldsymbol{r} + \boldsymbol{v}_J^{\mathrm{T}}\boldsymbol{r}\right) \right\rangle \\ &= D\left(a_I(\boldsymbol{\varepsilon}_I)_\beta G_I^{\boldsymbol{\Omega}} + a_J(\boldsymbol{\varepsilon}_J)_\beta G_J^{\boldsymbol{\Omega}}\right) \sum_{i=0}^{\infty} \sum_{j=0}^{\infty} \sum_{k=0}^{\infty} a_I^{2i+k} a_J^{2j+k} \left(\boldsymbol{\varepsilon}_I \boldsymbol{\varepsilon}_J\right)^k \frac{p_{II}^i p_{JJ}^j p_{IJ}^k}{i!j!k!} \end{aligned} \tag{S31}$$

with

$$G_I^{\boldsymbol{\Omega}} = \begin{cases} \displaystyle\sum_n \frac{q_n(\boldsymbol{A}_J \boldsymbol{A}_{IJ}^{-1} \boldsymbol{v}_I)_n}{m_n(E_i - E_f)} & \text{if } \boldsymbol{\Omega} = \boldsymbol{\mu}_{\mathrm{if}}^{(\mathrm{v})} \\ \displaystyle\sum_n q_n(\boldsymbol{A}_{IJ}^{-1} \boldsymbol{v}_I)_n & \text{if } \boldsymbol{\Omega} = \boldsymbol{\mu}_{\mathrm{if}}^{(\mathrm{l})} \end{cases}$$

$$G_J^{\boldsymbol{\Omega}} = \begin{cases} \displaystyle\sum_n \frac{q_n(\boldsymbol{A}_J \boldsymbol{A}_{IJ}^{-1} \boldsymbol{v}_J - \boldsymbol{v}_J)_n}{m_n(E_i - E_f)} & \text{if } \boldsymbol{\Omega} = \boldsymbol{\mu}_{\mathrm{if}}^{(\mathrm{v})} \\ \displaystyle\sum_n q_n(\boldsymbol{A}_{IJ}^{-1} \boldsymbol{v}_J)_n & \text{if } \boldsymbol{\Omega} = \boldsymbol{\mu}_{\mathrm{if}}^{(\mathrm{l})} \end{cases}$$

Step 2: Evaluate the derivatives at $a_I = a_J = 0$ of Eq. (S28):

$$\begin{aligned} I_2(\varepsilon_I, \varepsilon_J) &= \left. \frac{\partial^{2K_I + L_I}}{\partial a_I^{2K_I + L_I}} \frac{\partial^{2K_J + L_J}}{\partial a_J^{2K_J + L_J}} I_1(a_I, a_J) \right|_{a_I = a_J = 0} \\ &= D \sum_{i=0}^{\infty} \sum_{j=0}^{\infty} \sum_{k=0}^{\infty} \frac{p_{II}^i p_{JJ}^j p_{IJ}^k}{i!j!k!} \left(\boldsymbol{\varepsilon}_I^{\mathrm{T}} \boldsymbol{\varepsilon}_J\right)^k \kappa_I! \kappa_J! \left(G_I^{\boldsymbol{\Omega}}(\boldsymbol{\varepsilon}_I)_\beta \delta_{\kappa_I, 2i+k+1} \delta_{\kappa_J, 2j+k} + G_J^{\boldsymbol{\Omega}}(\boldsymbol{\varepsilon}_J)_\beta \delta_{\kappa_I, 2i+k} \delta_{\kappa_J, 2j+k+1}\right) \, . \end{aligned} \tag{S32}$$

## 3 Angular Integrals

We continue now with the third step in the evaluation scheme. The angular integral are derived separately for the components of $\alpha$ and then collected to form general expressions. In this part, the following relations and short-hand notations are used.

7

Associated Legendre Polynomials $\mathrm{P}_M^L(x)$:

$$\int_{-1}^{+1} \mathrm{P}_M^L(x)\,\mathrm{P}_M^{L'}(x)\,dx = \frac{2(L+M)!}{(2L+1)(L_M)!}\delta_{LL'} \tag{S33}$$

$$x\mathrm{P}_M^L(x) = \frac{(L-M+1)}{(2L+1)}\mathrm{P}_M^{L+1}(x) + \frac{(L+M)}{(2L+1)}\mathrm{P}_M^{L-1}(x) \tag{S34}$$

$$\sqrt{1-x^2}\mathrm{P}_M^L(x) = \frac{1}{2L+1}\left(\mathrm{P}_{M+1}^{L-1}(x) - \mathrm{P}_{M+1}^{L+1}(x)\right) \tag{S35}$$

Spherical Harmonics $\mathrm{Y}_M^L(\theta, \varphi)$:

$$\mathrm{Y}_M^L(\theta,\varphi) = N_L^M \exp(iM\varphi)\mathrm{P}_M^L(\cos(\theta)) \tag{S36}$$

$$N_L^M = \left(\frac{(2L+1)(L-M)!}{4\pi(L+M)!}\right)^{1/2} \tag{S37}$$

$$\int_0^\pi d\theta \sin(\theta) \int_0^{2\pi} d\varphi \left(\mathrm{Y}_M^L(\theta,\varphi)\right)^* \mathrm{Y}_{M'}^{L'}(\theta,\varphi) = \delta_{LL'}\delta_{MM'} \tag{S38}$$

$$\left(\boldsymbol{\varepsilon}_I^\mathrm{T}\boldsymbol{\varepsilon}_J\right)^k = \left(\boldsymbol{\varepsilon}_J^\mathrm{T}\boldsymbol{\varepsilon}_I\right)^k = \sum_{l=0}^{k} \mathrm{B}_{\frac{k-l}{2},l} \sum_{m=-l}^{l} \left(\mathrm{Y}_m^l(\hat{\varepsilon}_J)\right)^* \mathrm{Y}_m^l(\hat{\varepsilon}_I) \tag{S39}$$

with

$$B_{m,L} = \frac{8\pi(L+2m)!(L+m+1)!2^L}{m!(2L+2m+2)!} \quad m,L \in \mathbb{N}_0 \tag{S40}$$

## 3.1 (z)-Component

$$\int\int d\hat{\varepsilon}_I d\hat{\varepsilon}_J (\varepsilon_I)_z (\varepsilon_I^T \varepsilon_J)^k \left(Y_{M_1}^{L_1}(\hat{\varepsilon_I})\right)^* Y_{M_2}^{L_2}(\hat{\varepsilon_J})$$

$$= \sum_{l=0}^{k} \sum_{m=-l}^{l} \int\int d\hat{\varepsilon}_I d\hat{\varepsilon}_J B_{\frac{k-l}{2},l}(\varepsilon_I)_z \left(Y_{M_1}^{L_1}(\hat{\varepsilon_I})\right)^* Y_k^l(\hat{\varepsilon_I}) \left(Y_k^l(\hat{\varepsilon_J})\right)^* Y_{M_2}^{L_2}(\hat{\varepsilon_J})$$

$$= B_{\frac{k-L_2}{2},L_2} \int d\varepsilon_I (\varepsilon_I)_z \left(Y_{M_1}^{L_1}(\hat{\varepsilon_I})\right)^* Y_{M_2}^{L_2}(\hat{\varepsilon_I})$$

$$= B_{\frac{k-L_2}{2},L_2} N_{L_1}^{M_1} N_{L_2}^{M_2} \int_0^{\pi} \int_0^{2\pi} d\theta d\varphi \ \sin(\theta)\cos(\theta)\exp\left(i(M_2-M_1)\varphi\right) P_{M_1}^{L_1}(\cos(\theta)) P_{M_2}^{L_2}(\cos(\theta))$$

$$= 2\pi B_{\frac{k-L_2}{2},L_2} N_{L_1}^{M_1} N_{L_2}^{M_2} \delta_{M_1,M_2} \int_0^{\pi} d\theta \ \sin(\theta)\cos(\theta) P_{M_1}^{L_1}(\cos(\theta)) P_{M_2}^{L_2}(\cos(\theta))$$

$$= 2\pi B_{\frac{k-L_2}{2},L_2} N_{L_1}^{M_1} N_{L_2}^{M_2} \delta_{M_1,M_2} \int_{-1}^{+1} dx \ x P_{M_1}^{L_1}(x) P_{M_2}^{L_2}(x)$$

$$= 2\pi B_{\frac{k-L_2}{2},L_2} N_{L_1}^{M_1} N_{L_2}^{M_2} \left(\frac{L_2-M_2+1}{2L_2+1}\right) \delta_{M_1,M_2} \int_{-1}^{+1} dx \ P_M^{L_1}(x) P_M^{L_2+1}(x)$$

$$+ 2\pi B_{\frac{k-L_2}{2},L_2} N_{L_1}^{M_1} N_{L_2}^{M_2} \left(\frac{L_2+M_2}{2L_2+1}\right) \delta_{M_1,M_2} \int_{-1}^{+1} dx \ P_M^{L_1}(x) P_M^{L_2-1}(x)$$

$$= 2\pi B_{\frac{k-L_2}{2},L_2} N_{L_1}^{M_1} N_{L_2}^{M_2} \left(\frac{L_2-M_2+1}{2L_2+1}\right) \left(\frac{2(L_1+M)!}{(2L_1+1)(L_1-M)!}\right) \delta_{L_1,L_2+1} \delta_{M_1,M_2}$$

$$+ 2\pi B_{\frac{k-L_2}{2},L_2} N_{L_1}^{M_1} N_{L_2}^{M_2} \left(\frac{L_2+M_2}{2L_2+1}\right) \left(\frac{2(L_1+M_1)!}{(2L_1+1)(L_1-M_1)!}\right) \delta_{L_1+1,L_2} \delta_{M_1,M_2}$$

$$= B_{\frac{k-L_2}{2},L_2} \left(\frac{(L_2+M_2+1)(L_2-M_2+1)}{(2L_2+1)(2L_2+3)}\right)^{1/2} \delta_{L_1,L_2+1} \delta_{M_1,M_2}$$

$$+ B_{\frac{k-L_2}{2},L_2} \left(\frac{(L_1+M_1+1)(L_1-M_1+1)}{(2L_1+1)(2L_1+3)}\right)^{1/2} \delta_{L_1+1,L_2} \delta_{M_1,M_2}$$

$$= B_{\frac{k-L_2}{2},L_2} C_1^z(L_2,M_1,M_2) \delta_{L_1,L_2+1} + B_{\frac{k-L_2}{2},L_2} C_2^z(L_1,M_1,M_2) \delta_{L_1+1,L_2} \quad (S41)$$

where

$$C_1^z(L_2,M_2,M_1) = \left(\frac{(L_2+M_2+1)(L_2-M_2+1)}{(2L_2+1)(2L_2+3)}\right)^{1/2} \delta_{M_1,M_2} \quad (S42)$$

$$C_2^z(L_1,M_1,M_2) = \left(\frac{(L_1+M_1+1)(L_1-M_1+1)}{(2L_1+1)(2L_1+3)}\right)^{1/2} \delta_{M_1,M_2} \quad (S43)$$

Analogously for

$$\int\int d\hat{\varepsilon}_I d\hat{\varepsilon}_J (\varepsilon_J)_z (\varepsilon_I^T \varepsilon_J)^k \left(Y_{M_1}^{L_1}(\hat{\varepsilon_I})\right)^* Y_{M_2}^{L_2}(\hat{\varepsilon_J})$$

$$= B_{\frac{k-L_1}{2},L_1} C_1^z(L_2,M_2,M_1) \delta_{L_1,L_2+1} + B_{\frac{k-L_1}{2},L_1} C_2^z(L_1,M_1,M_2) \delta_{L_1+1,L_2} \quad (S44)$$

## 3.2 (−)-Component

$$\int\int d\hat{\varepsilon}_I d\hat{\varepsilon}_J \left[(\varepsilon_I)_x + i(\varepsilon_I)_y\right] (\varepsilon_I{}^T \varepsilon_J)^k \left(Y_{M_1}^{L_1}(\hat{\varepsilon}_I)\right)^* Y_{M_2}^{L_2}(\hat{\varepsilon}_J)$$

$$= B_{\frac{k-L_2}{2}, L_2} \int d\hat{\varepsilon}_I \left[(\varepsilon_I)_x + i(\varepsilon_I)_y\right] \left(Y_{M_1}^{L_1}(\hat{\varepsilon}_I)\right)^* Y_{M_2}^{L_2}(\hat{\varepsilon}_I)$$

$$= B_{\frac{k-L_2}{2}, L_2} N_{L_1}^{M_1} N_{L_2}^{M_2} \int_0^{2\pi} d\varphi \ [\cos(\varphi) + i\sin(\varphi)] \exp(i(M_2 - M_1)\varphi) \int_0^{\pi} d\theta \ \sin(\theta)^2 P_{M_1}^{L_1}(\cos(\theta)) P_{M_2}^{L_2}(\cos(\theta))$$

$$= B_{\frac{k-L_2}{2}, L_2} N_{L_1}^{M_1} N_{L_2}^{M_2} \int_0^{2\pi} d\varphi \ \exp(i(M_2 + 1 - M_1)\varphi) \int_{-1}^{+1} dx \ \sqrt{1-x^2} P_{M_1}^{L_1}(x) P_{M_2}^{L_2}(x)$$

$$= B_{\frac{k-L_2}{2}, L_2} N_{L_1}^{M_1} N_{L_2}^{M_2} 2\pi \delta_{M_1, M_2+1} \int_{-1}^{+1} dx \ \sqrt{1-x^2} P_{M_1}^{L_1}(x) P_{M_2}^{L_2}(x)$$

$$= B_{\frac{k-L_2}{2}, L_2} N_{L_1}^{M_1} N_{L_2}^{M_2} \frac{2\pi}{2L_2+1} \delta_{M_1, M_2+1} \int_{-1}^{+1} dx \ P_{M_2+1}^{L_1}(x) P_{M_2+1}^{L_2-1}(x) - P_{M_2+1}^{L_1}(x) P_{M_2+1}^{L_2+1}(x)$$

$$= 4\pi B_{\frac{k-L_2}{2}, L_2} N_{L_1}^{M_1} N_{L_2}^{M_2} \frac{(L_1+M_1)!}{(2L_2+1)(2L_1+1)(L_1-M_1)!} \delta_{M_1, M_2+1} \left(\delta_{L_1+1, L_2} - \delta_{L_1, L_2+1}\right)$$

$$= B_{\frac{k-L_2}{2}, L_2} \delta_{M_1, M_2+1} \left(\left(\frac{(L_1-M_1+2)(L_1-M_1+1)}{(2L_1+1)(2L_1+3)}\right)^{1/2} \delta_{L_1+1, L_2} - \left(\frac{(L_2+M_2+2)(L_2+M_2+1)}{(2L_2+1)(2L_2+3)}\right)^{1/2} \delta_{L_1, L_2+1}\right)$$

$$= B_{\frac{k-L_2}{2}, L_2} C_1^-(L_2, M_1, M_2) \delta_{L_1, L_2+1} + B_{\frac{k-L_2}{2}, L_2} C_2^-(L_1, M_1, M_2) \delta_{L_1+1, L_2} \quad \text{(S45)}$$

where

$$C_1^-(L_2, M_2, M_1) = -\left(\frac{(L_2+M_2+2)(L_2+M_2+1)}{(2L_2+1)(2L_2+3)}\right)^{1/2} \delta_{M_1, M_2+1} \quad \text{(S46)}$$

$$C_2^-(L_1, M_1, M_2) = \left(\frac{(L_1-M_1+2)(L_1-M_1+1)}{(2L_1+1)(2L_1+3)}\right)^{1/2} \delta_{M_1, M_2+1} \quad \text{(S47)}$$

Analogously for

$$\int\int d\hat{\varepsilon}_I d\hat{\varepsilon}_J \left[(\varepsilon_J)_x + i(\varepsilon_J)_y\right] (\varepsilon_I^T \varepsilon_J)^k \left(Y_{M_1}^{L_1}(\hat{\varepsilon}_I)\right)^* Y_{M_2}^{L_2}(\hat{\varepsilon}_J)$$

$$= B_{\frac{k-L_1}{2}, L_1} C_1^-(L_2, M_2, M_1) \delta_{L_1, L_2+1} + B_{\frac{k-L_1}{2}, L_1} C_2^-(L_1, M_1, M_2) \delta_{L_1+1, L_2} \quad \text{(S48)}$$

## 3.3 (+)-Component

$$\int\int d\hat{\varepsilon}_I d\hat{\varepsilon}_J \left[(\varepsilon_I)_x - i(\varepsilon_I)_y\right] (\varepsilon_I{}^T\varepsilon_J)^k \left(Y_{M_1}^{L_1}(\hat{\varepsilon}_I)\right)^* Y_{M_2}^{L_2}(\hat{\varepsilon}_J)$$

$$= B_{\frac{k-L_2}{2},L_2} \int d\hat{\varepsilon}_I \left[(\varepsilon_I)_x - i(\varepsilon_I)_y\right] \left(Y_{M_1}^{L_1}(\hat{\varepsilon}_I)\right)^* Y_{M_2}^{L_2}(\hat{\varepsilon}_I)$$

$$= B_{\frac{k-L_2}{2},L_2} N_{L_1}^{M_1} N_{L_2}^{M_2} \int_0^{2\pi} d\varphi \; [\cos(\varphi) - i\sin(\varphi)] \exp(i(M_2 - M_1)\varphi) \int_0^\pi d\theta \; \sin(\theta)^2 P_{M_1}^{L_1}(\cos(\theta)) P_{M_2}^{L_2}(\cos(\theta))$$

$$= B_{\frac{k-L_2}{2},L_2} N_{L_1}^{M_1} N_{L_2}^{M_2} \int_0^\pi d\varphi \exp(i(M_2 - M_1 - 1)\varphi) \int_{-1}^{+1} dx \; \sqrt{1-x^2} P_{M_1}^{L_1}(x) P_{M_2}^{L_2}(x)$$

$$= B_{\frac{k-L_2}{2},L_2} N_{L_1}^{M_1} N_{L_2}^{M_2} 2\pi \delta_{M_1+1,M_2} \int_{-1}^{+1} dx \; \sqrt{1-x^2} P_{M_1}^{L_1}(x) P_{M_2}^{L_2}(x)$$

$$= B_{\frac{k-L_2}{2},L_2} N_{L_1}^{M_1} N_{L_2}^{M_2} \frac{2\pi}{2L_1 + 1} \delta_{M_1+1,M_2} \int_{-1}^{+1} dx \; P_{M_1+1}^{L_1-1}(x) P_{M_1+1}^{L_2}(x) - P_{M_1+1}^{L_1+1}(x) P_{M_1+1}^{L_2}(x)$$

$$= 4\pi B_{\frac{k-L_2}{2},L_2} N_{L_1}^{M_1} N_{L_2}^{M_2} \frac{(L_2 + M_2)!}{(2L_1+1)(2L_2+1)(L_2-M_2)!} \delta_{M_1+1,M_2} \left(\delta_{L_1,L_2+1} - \delta_{L_1+1,L_2}\right)$$

$$= B_{\frac{k-L_2}{2},L_2} \delta_{M_1+1,M_2} \left(\left(\frac{(L_2-M_2+2)(L_2-M_2+1)}{(2L_2+1)(2L_2+3)}\right)^{1/2} \delta_{L_1,L_2+1} - \left(\frac{(L_1+M_1+2)(L_1+M_1+1)}{(2L_1+1)(2L_1+3)}\right)^{1/2} \delta_{L_1+1,L_2}\right)$$

$$= B_{\frac{k-L_2}{2},L_2} C_1^+(L_2, M_2, M_1) \delta_{L_1,L_2+1} + B_{\frac{k-L_2}{2},L_2} C_2^+(L_1, M_1, M_2) \delta_{L_1+1,L_2} \tag{S49}$$

where

$$C_1^+(L_2, M_2, M_1) = \left(\frac{(L_2-M_2+2)(L_2-M_2+1)}{(2L_2+1)(2L_2+3)}\right)^{1/2} \delta_{M_1+1,M_2} \tag{S50}$$

$$C_2^+(L_1, M_1, M_2) = -\left(\frac{(L_1+M_1+2)(L_1+M_1+1)}{(2L_1+1)(2L_1+3)}\right)^{1/2} \delta_{M_1+1,M_2} \tag{S51}$$

Analogously for

$$\int\int d\hat{\varepsilon}_I d\hat{\varepsilon}_J \left[(\varepsilon_J)_x + i(\varepsilon_J)_y\right] (\varepsilon_I{}^T\varepsilon_J)^k \left(Y_{M_1}^{L_1}(\hat{\varepsilon}_I)\right)^* Y_{M_2}^{L_2}(\hat{\varepsilon}_J)$$

$$= B_{\frac{k-L_1}{2},L_1} C_1^+(L_2, M_2, M_1) \delta_{L_1,L_2+1} + B_{\frac{k-L_1}{2},L_1} C_2^+(L_1, M_1, M_2) \delta_{L_1+1,L_2} \tag{S52}$$

## 3.4 General Expressions

We can now collect Eqs. (S41), (S45) and (S49) and generalize them as:

$$\int\int d\hat{\varepsilon}_I d\hat{\varepsilon}_J (\varepsilon_I)_\beta (\varepsilon_I{}^T\varepsilon_J)^k \left(Y_{M_1}^{L_1}(\hat{\varepsilon}_I)\right)^* Y_{M_2}^{L_2}(\hat{\varepsilon}_J)$$

$$= B_{\frac{k-L_2}{2},L_2} C_1^\alpha(L_2, M_2, M_1) \delta_{L_1,L_2+1} + B_{\frac{k-L_2}{2},L_2} C_2^\alpha(L_1, M_1, M_2) \delta_{L_1+1,L_2} \tag{S53}$$

and

$$\int\int d\hat{\varepsilon}_I d\hat{\varepsilon}_J (\varepsilon_J)_\beta (\varepsilon_I{}^T\varepsilon_J)^k \left(Y_{M_1}^{L_1}(\hat{\varepsilon}_I)\right)^* Y_{M_2}^{L_2}(\hat{\varepsilon}_J)$$

$$= B_{\frac{k-L_1}{2},L_1} C_1^\alpha(L_2, M_2, M_1) \delta_{L_1,L_2+1} + B_{\frac{k-L_1}{2},L_1} C_2^\alpha(L_1, M_1, M_2) \delta_{L_1+1,L_2} \tag{S54}$$

with

$$C_1^\alpha(L_2, M_2, M_1) = \begin{cases} \left(\frac{(L_2+M_2+1)(L_2-M_2+1)}{(2L_2+1)(2L_2+3)}\right)^{1/2} \delta_{M_1,M_2} & \text{if } \alpha = z \quad \text{(Eq. (S42))} \\ \left(\frac{(L_2-M_2+2)(L_2-M_2+1)}{(2L_2+1)(2L_2+3)}\right)^{1/2} \delta_{M_1+1,M_2} & \text{if } \alpha = + \quad \text{(Eq. (S50))} \\ -\left(\frac{(L_2+M_2+2)(L_2+M_2+1)}{(2L_2+1)(2L_2+3)}\right)^{1/2} \delta_{M_1,M_2+1} & \text{if } \alpha = - \quad \text{(Eq. (S46))} \end{cases} \tag{S55}$$

and

$$C_2^\alpha(L_1, M_1, M_2) = \begin{cases} \left(\frac{(L_1+M_1+1)(L_1-M_1+1)}{(2L_1+1)(2L_1+3)}\right)^{1/2} \delta_{M_1,M_2} & \text{if } \alpha = z \quad \text{(Eq. (S43))} \\ -\left(\frac{(L_1+M_1+2)(L_1+M_1+1)}{(2L_1+1)(2L_1+3)}\right)^{1/2} \delta_{M_1+1,M_2} & \text{if } \alpha = + \quad \text{(Eq. (S51))} \\ \left(\frac{(L_1-M_1+2)(L_1-M_1+1)}{(2L_1+1)(2L_1+3)}\right)^{1/2} \delta_{M_1,M_2+1} & \text{if } \alpha = - \quad \text{(Eq. (S47))} \end{cases} \tag{S56}$$

Using Eqs. (S53) and (S54) for the angular integrals and Eq. (S32) as $I_2(\varepsilon_I, \varepsilon_J)$ we find the final result $I_3$ of step 3:

$$\begin{aligned} I_3 &= \frac{1}{B_{K_I L_I} B_{K_J L_J}} \int d\hat{\varepsilon}_I \int d\hat{\varepsilon}_J Y_{M_{LI}}^{L_I *}(\hat{\varepsilon}_I) Y_{M_{LJ}}^{L_J}(\hat{\varepsilon}_J) I_2(\varepsilon_I, \varepsilon_J) \\ &= \frac{\kappa_I! \kappa_J!}{B_{K_I L_1} B_{K_J L_2}} D \sum_{i=0}^{\infty} \sum_{j=0}^{\infty} \sum_{k=0}^{\infty} \frac{p_{II}^i p_{JJ}^j p_{IJ}^k}{i! j! k!} C_1^\alpha(L_2, M_2, M_1) \\ &\times \left( G_I^\Omega B_{\frac{k-L_2}{2}, L_2} \delta_{\kappa_I, 2i+k+1} \delta_{\kappa_J, 2j+k} + G_J^\Omega B_{\frac{k-L_1}{2}, L_1} \delta_{\kappa_I, 2i+k} \delta_{\kappa_J, 2j+k+1} \right) \end{aligned} \tag{S57}$$

# 4 Final Expressions

In this part, we collect the results from sections 2 and 3 to form the final integral expressions.

## 4.1 Summation Indices

In order to limit the summation indices we exploit the properties of the $B_{\frac{m-L}{2}, L}$ factor described in Eq. (S57): We require that $\frac{k-L}{2} \in \mathbb{N}_0$.

### 4.1.1 Case $L_1 = L_2 + 1$

$$I_3 = \frac{\kappa_I! \kappa_J!}{B_{K_I L_1} B_{K_J L_2}} D \sum_{i=0}^{\infty} \sum_{j=0}^{\infty} \sum_{k=0}^{\infty} \frac{p_{II}^i p_{JJ}^j p_{IJ}^k}{i! j! k!} C_1^\alpha(L_2, M_2, M_1)$$

$$\times \left( G_I^{\boldsymbol{\Omega}} \mathrm{B}_{\frac{k-L_2}{2}, L_2} \delta_{\kappa_I, 2i+k+1} \delta_{\kappa_J, 2j+k} + G_J^{\boldsymbol{\Omega}} \mathrm{B}_{\frac{k-L_1}{2}, L_1} \delta_{\kappa_I, 2i+k} \delta_{\kappa_J, 2j+k+1} \right)$$

$$= \frac{\kappa_I! \kappa_J!}{B_{K_I L_1} B_{K_J L_2}} D \sum_{i=0}^{\infty} \sum_{j=0}^{\infty} \sum_{k=0}^{\infty} \frac{p_{II}^i p_{JJ}^j p_{IJ}^k}{i! j! k!} C_1^\alpha(L_2, M_2, M_1) G_I^{\boldsymbol{\Omega}} \mathrm{B}_{\frac{k-L_2}{2}, L_2} \delta_{\kappa_I, 2i+k+1} \delta_{\kappa_J, 2j+k} \quad \text{(S58)}$$

$$+ \frac{\kappa_I! \kappa_J!}{B_{K_I L_1} B_{K_J L_2}} D \sum_{i=0}^{\infty} \sum_{j=0}^{\infty} \sum_{k=0}^{\infty} \frac{p_{II}^i p_{JJ}^j p_{IJ}^k}{i! j! k!} C_1^\alpha(L_2, M_2, M_1) G_J^{\boldsymbol{\Omega}} \mathrm{B}_{\frac{k-L_1}{2}, L_1} \delta_{\kappa_I, 2i+k} \delta_{\kappa_J, 2j+k+1} \quad \text{(S59)}$$

Summation indices for term (S58)

$$\frac{k - L_2}{2} = m \quad \Rightarrow \quad k = 2m + L_2 \tag{S60}$$

$$\kappa_I = 2K_I + L_1 = 2K_I + L_2 + 1 = 2i + 2m + L_2 + 1 \quad \Rightarrow \quad i = K_I - m \tag{S61}$$

$$\kappa_J = 2K_J + L_2 = 2j + 2m + L_2 \quad \Rightarrow \quad j = K_J - m \tag{S62}$$

Summation indices for term (S59)

$$\frac{k - L_1}{2} = m - 1 \quad \Rightarrow \quad k = 2m + L_1 - 2 \tag{S63}$$

$$\kappa_I = 2K_I + L_1 = 2i + 2m - 2 + L_1 \quad \Rightarrow \quad i = K_I - m + 1 \tag{S64}$$

$$\kappa_J = 2K_J + L_2 = 2j + 2m + L_1 - 1 = 2j + 2m + L_2 \quad \Rightarrow \quad j = K_J - m \tag{S65}$$

Insertion of these new summation indices into terms (S58) and (S59) leads to

$$I_3 = \frac{\kappa_I! \kappa_J!}{B_{K_I L_1} B_{K_J L_2}} D p_{II}^{K_I} p_{JJ}^{K_J} p_{IJ}^{L_2} G_I^{\boldsymbol{\Omega}} \sum_{m=0}^{\min(K_I, K_J)} \left( \frac{p_{IJ} p_{IJ}}{p_{II} p_{JJ}} \right)^m \left( \frac{C_1^\alpha(L_2, M_2, L_1) B_{m L_2}}{(K_I - m)!(K_J - m)!(2m + L_2)!} \right)$$

$$+ \frac{\kappa_I! \kappa_J!}{B_{K_I L_1} B_{K_J L_2}} D p_{II}^{K_I + 1} p_{JJ}^{K_J} p_{IJ}^{L_2 - 1} G_J^{\boldsymbol{\Omega}} \sum_{m=1}^{\min(K_I + 1, K_J)} \left( \frac{p_{IJ} p_{IJ}}{p_{II} p_{JJ}} \right)^m \left( \frac{C_1^\alpha(L_2, M_2, M_1) B_{(m-1) L_1}}{(K_I - m + 1)!(K_J - m)!(2m + L_1 - 2)!} \right)$$

$$= \frac{\kappa_I! \kappa_J!}{B_{K_I L_1} B_{K_J L_2}} D p_{II}^{K_I} p_{JJ}^{K_J} p_{IJ}^{L_2} C_1^\alpha(L_2, M_2, M_1) \times$$

$$\left( G_I^{\boldsymbol{\Omega}} \sum_{m=0}^{\min(K_I, K_J)} \left( \frac{p_{IJ} p_{IJ}}{p_{II} p_{JJ}} \right)^m \left( \frac{B_{m L_2}}{(K_I - m)!(K_J - m)!(2m + L_2)!} \right) \right.$$

$$\left. + \frac{G_J^{\boldsymbol{\Omega}} p_{II}}{p_{IJ}} \sum_{m=1}^{\min(K_I + 1, K_J)} \left( \frac{p_{IJ} p_{IJ}}{p_{II} p_{JJ}} \right)^m \left( \frac{B_{(m-1) L_1}}{(K_I - m + 1)!(K_J - m)!(2m + L_1 - 2)!} \right) \right) \tag{S66}$$

### 4.1.2 Case $L_1 + 1 = L_2$

$$I_3 = \frac{\kappa_I!\kappa_J!}{B_{K_IL_1}B_{K_JL_2}}D\sum_{i=0}^{\infty}\sum_{j=0}^{\infty}\sum_{k=0}^{\infty}\frac{p_{II}^i p_{JJ}^j p_{IJ}^k}{i!j!k!}C_2^\alpha(L_1, M_1, M_2)$$
$$\times \left(G_I^{\mathbf{\Omega}}\mathrm{B}_{\frac{k-L_2}{2},L_2}\delta_{\kappa_I,2i+k+1}\delta_{\kappa_J,2j+k} + G_J^{\mathbf{\Omega}}\mathrm{B}_{\frac{k-L_1}{2},L_1}\delta_{\kappa_I,2i+k}\delta_{\kappa_J,2j+k+1}\right)$$
$$= \frac{\kappa_I!\kappa_J!}{B_{K_IL_1}B_{K_JL_2}}D\sum_{i=0}^{\infty}\sum_{j=0}^{\infty}\sum_{k=0}^{\infty}\frac{p_{II}^i p_{JJ}^j p_{IJ}^k}{i!j!k!}C_2^\alpha(L_1, M_1, M_2)G_I^{\mathbf{\Omega}}\mathrm{B}_{\frac{k-L_2}{2},L_2}\delta_{\kappa_I,2i+k+1}\delta_{\kappa_J,2j+k} \quad (S67)$$
$$+ \frac{\kappa_I!\kappa_J!}{B_{K_IL_1}B_{K_JL_2}}D\sum_{i=0}^{\infty}\sum_{j=0}^{\infty}\sum_{k=0}^{\infty}\frac{p_{II}^i p_{JJ}^j p_{IJ}^k}{i!j!k!}C_2^\alpha(L_1, M_1, M_2)G_J^{\mathbf{\Omega}}\mathrm{B}_{\frac{k-L_1}{2},L_1}\delta_{\kappa_I,2i+k}\delta_{\kappa_J,2j+k+1} \quad (S68)$$

Summation indices for term (S67)

$$\frac{k-L_2}{2} = m - 1 \quad \Rightarrow \quad k = 2m + L_2 - 2 \tag{S69}$$

$$\kappa_I = 2K_I + L_1 = 2i + 2m + L_2 - 1 = 2i + 2m + L_1 \quad \Rightarrow \quad i = K_I - m \tag{S70}$$

$$\kappa_J = 2K_J + L_2 = 2j + 2m + L_2 - 2 \quad \Rightarrow \quad j = K_J - m + 1 \tag{S71}$$

Summation indices for term (S68)

$$\frac{k-L_1}{2} = m \quad \Rightarrow \quad k = 2m + L_1 \tag{S72}$$

$$\kappa_I = 2K_I + L_1 = 2i + 2m + L_1 \quad \Rightarrow \quad i = K_I - m \tag{S73}$$

$$\kappa_J = 2K_J + L_2 = 2j + 2m + L_1 + 1 = 2j + 2m + L_2 \quad \Rightarrow \quad j = K_J - m \tag{S74}$$

Insertion of these new summation indices into terms (S67) and (S68) leads to

$$I_3 = \frac{\kappa_I!\kappa_J!}{B_{K_IL_1}B_{K_JL_2}}Dp_{II}^{K_I}p_{JJ}^{K_J+1}p_{IJ}^{L_1-1}G_I^{\mathbf{\Omega}}\sum_{m=1}^{\min(K_I,K_J+1)}\left(\frac{p_{IJ}p_{IJ}}{p_{II}p_{JJ}}\right)^m\left(\frac{C_2^\alpha(L_1,M_1,M_2)B_{(m-1)L_2}}{(K_I-m)!(K_J-m+1)!(2m+L_2-2)!}\right)$$
$$+\frac{\kappa_I!\kappa_J!}{B_{K_IL_1}B_{K_JL_2}}Dp_{II}^{K_I}p_{JJ}^{K_J}p_{IJ}^{L_1}G_J^{\mathbf{\Omega}}\sum_{m=0}^{\min(K_I,K_J)}\left(\frac{p_{IJ}p_{IJ}}{p_{II}p_{JJ}}\right)^m\left(\frac{C_2^\alpha(L_1,M_1,M_2)B_{mL_1}}{(K_I-m)!(K_J-m)!(2m+L_1)!}\right)$$
$$=\frac{\kappa_I!\kappa_J!}{B_{K_IL_1}B_{K_JL_2}}Dp_{II}^{K_I}p_{JJ}^{K_J}p_{IJ}^{L_1}C_2^\alpha(L_1,M_1,M_2)\times$$
$$\left(\frac{G_I^{\mathbf{\Omega}}p_{JJ}}{p_{IJ}}\sum_{m=1}^{\min(K_I,K_J+1)}\left(\frac{p_{IJ}p_{IJ}}{p_{II}p_{JJ}}\right)^m\left(\frac{B_{(m-1)L_2}}{(K_I-m)!(K_J-m+1)!(2m+L_2-2)!}\right)\right.$$
$$\left.+G_J^{\mathbf{\Omega}}\sum_{m=0}^{\min(K_I,K_J)}\left(\frac{p_{IJ}p_{IJ}}{p_{II}p_{JJ}}\right)^m\left(\frac{B_{mL_1}}{(K_I-m)!(K_J-m)!(2m+L_1)!}\right)\right) \tag{S75}$$

## 4.2 Quasi-Normalisation Factor

We quasi normalize our basis functions according to Eq. (TK) in the original paper as:

$$\left\langle \phi_I^{(L_1,M_1)} \middle| \phi_I^{(L_1,M_1)} \right\rangle = \left\langle \exp\left(-\frac{1}{2}\boldsymbol{r}^{\mathrm{T}}(\boldsymbol{A}_I \otimes \boldsymbol{1}_3)\boldsymbol{r} + \boldsymbol{v}_I^{\mathrm{T}}\boldsymbol{r}\right) \middle| \exp\left(-\frac{1}{2}\boldsymbol{r}^{\mathrm{T}}(\boldsymbol{A}_I \otimes \boldsymbol{1}_3)\boldsymbol{r} + \boldsymbol{v}_I^{\mathrm{T}}\boldsymbol{r}\right) \right\rangle$$
$$= \frac{\kappa_I!\kappa_I!}{B_{K_IL_1}B_{K_IL_1}} \left(\frac{(2\pi)^N}{\det(2\boldsymbol{A}_I)}\right)^{3/2} q_I^{2K_I+L_1} F(K_I, L_1) \tag{S76}$$

with

$$F(K, L) = \sum_{m=0}^{K} \frac{4^m B_{mL}}{4^K (K-m)!(K-m)!(L+2m)!} \tag{S77}$$

$$q_I = \frac{1}{2}\boldsymbol{u}_I^{\mathrm{T}}\boldsymbol{A}_I^{-1}\boldsymbol{u}_I \tag{S78}$$

With this, the quasi-normalization factor of the integral becomes

$$\left|\phi_I^{(L_1,M_1)}\right|\left|\phi_J^{(L_2,M_2)}\right| = \frac{\kappa_I!\kappa_J!}{B_{K_IL_1}B_{K_JL_2}} \left(\frac{(2\pi)^{2N}}{\det(2\boldsymbol{A}_I)\det(2\boldsymbol{A}_J)}\right)^{3/4} q_I^{K_I+L_1/2} q_J^{K_J+L_2/2} \left(F(K_I,L_1)F(K_J,L_2)\right)^{1/2} \tag{S79}$$

## 4.3 Quasi-Normalised Integral Expressions

We now form the quasi normalized integral expressions from Eq. (S79) and Eq. (S66) for $L_1 = L_2 + 1$ and Eq. (S75) $L_1 + 1 = L_2$ as

$$\frac{\left\langle \phi_I^{(L_1,M_1)} \middle| \boldsymbol{\Omega}_\alpha \middle| \phi_J^{(L_2,M_2)} \right\rangle}{\left|\phi_I^{(L_1,M_1)}\right|\left|\phi_J^{(L_2,M_2)}\right|} = \frac{I_3}{\left|\phi_I^{(L_1,M_1)}\right|\left|\phi_J^{(L_2,M_2)}\right|} \tag{S80}$$

### 4.3.1 Case $L_1 = L_2 + 1$

$$\frac{\left\langle \phi_I^{(L_1,M_1)} \middle| \mathbf{\Omega}_\alpha \middle| \phi_J^{(L_2,M_2)} \right\rangle}{\left| \phi_I^{(L_1,M_1)} \right| \left| \phi_J^{(L_2,M_2)} \right|}$$

$$= \left( \frac{\det(2A_I)\det(2A_J)}{\det(A_{IJ})\det(A_{IJ})} \right)^{3/4} \left( \frac{p_{II}}{q_I} \right)^{K_I} \left( \frac{p_{JJ}}{q_J} \right)^{K_J} \left( \frac{p_{IJ}}{\sqrt{q_I q_J}} \right)^{L_2} C_1^\alpha(L_2, M_2, M_1) \left( q_I F(K_I, L_1) F(K_J, L_2) \right)^{-1/2} \times$$

$$\left[ G_I^{\mathbf{\Omega}} \sum_{m=0}^{\min(K_I, K_J)} \left( \frac{p_{IJ} p_{IJ}}{p_{II} p_{JJ}} \right)^m \left( \frac{B_{mL_2}}{(K_I - m)!(K_J - m)!(2m + L_2)!} \right) \right.$$

$$\left. + \frac{G_J^{\mathbf{\Omega}} p_{II}}{p_{IJ}} \sum_{m=1}^{\min(K_I+1, K_J)} \left( \frac{p_{IJ} p_{IJ}}{p_{II} p_{JJ}} \right)^m \left( \frac{B_{(m-1)L_1}}{(K_I - m + 1)!(K_J - m)!(2m + L_1 - 2)!} \right) \right]$$

$$= \left( \frac{\det(2A_I)\det(2A_J)}{\det(A_{IJ})\det(A_{IJ})} \right)^{3/4} \left( \frac{p_{II}}{q_I} \right)^{K_I} \left( \frac{p_{JJ}}{q_J} \right)^{K_J} \left( \frac{p_{IJ}}{\sqrt{q_I q_J}} \right)^{L_2} C_1^\alpha(L_2, M_2, M_1) \left( q_I \right)^{-1/2} \times$$

$$\left[ G_I^{\mathbf{\Omega}} \sum_{m=0}^{\min(K_I, K_J)} \left( \frac{p_{IJ} p_{IJ}}{p_{II} p_{JJ}} \right)^m H_1(m, K_I, K_J, L_2) + \frac{G_J^{\mathbf{\Omega}} p_{II}}{p_{IJ}} \sum_{m=1}^{\min(K_I+1, K_J)} \left( \frac{p_{IJ} p_{IJ}}{p_{II} p_{JJ}} \right)^m H_2(m, K_I, K_J, L_1) \right] \tag{S81}$$

with

$$p_{XY} = \frac{1}{2} \mathbf{v}_X \mathbf{A}_{IJ}^{-1} \mathbf{v}_Y \quad \text{with} \quad X, Y \in \{I, J\} \tag{S82}$$

$$q_X = \frac{1}{2} \mathbf{v}_X^\mathrm{T} \mathbf{A}_X^{-1} \mathbf{v}_X \quad \text{with} \quad X \in \{I, J\} \tag{S83}$$

$$H_1(m, K_1, K_2, L) = \frac{B_{mL}}{(K_1 - m)!(K_2 - m)!(2m + L)!} \left[ F(K_1, L+1) F(K_2, L) \right]^{-1/2} \tag{S84}$$

$$H_2(m, K_1, K_2, L) = \frac{B_{(m-1)L}}{(K_1 - m + 1)!(K_2 - m)!(2m + L - 2)!} \left[ F(K_1, L) F(K_2, L-1) \right]^{-1/2} \tag{S85}$$

$$G_I^{\mathbf{\Omega}} = \begin{cases} \displaystyle\sum_n \frac{q_n (\mathbf{A}_J \mathbf{A}_{IJ}^{-1} \mathbf{v}_I)_n}{m_n (E_i - E_f)} & \text{if } \mathbf{\Omega} = \boldsymbol{\mu}_{\text{if}}^{(\text{v})} \\ \displaystyle\sum_n q_n (\mathbf{A}_{IJ}^{-1} \mathbf{v}_I)_n & \text{if } \mathbf{\Omega} = \boldsymbol{\mu}_{\text{if}}^{(\text{l})} \end{cases}$$

$$G_J^{\mathbf{\Omega}} = \begin{cases} \displaystyle\sum_n \frac{q_n (\mathbf{A}_J \mathbf{A}_{IJ}^{-1} \mathbf{v}_J - \mathbf{v}_J)_n}{m_n (E_i - E_f)} & \text{if } \mathbf{\Omega} = \boldsymbol{\mu}_{\text{if}}^{(\text{v})} \\ \displaystyle\sum_n q_n (\mathbf{A}_{IJ}^{-1} \mathbf{v}_J)_n & \text{if } \mathbf{\Omega} = \boldsymbol{\mu}_{\text{if}}^{(\text{l})} \end{cases} \tag{S86}$$

$$C_1^\alpha(L_2, M_2, M_1) = \begin{cases} \left( \frac{(L_2 + M_2 + 1)(L_2 - M_2 + 1)}{(2L_2 + 1)(2L_2 + 3)} \right)^{1/2} \delta_{M_1, M_2} & \text{if } \alpha = z \quad (\text{Eq. (S42)}) \\ \left( \frac{(L_2 - M_2 + 2)(L_2 - M_2 + 1)}{(2L_2 + 1)(2L_2 + 3)} \right)^{1/2} \delta_{M_1+1, M_2} & \text{if } \alpha = + \quad (\text{Eq. (S50)}) \\ -\left( \frac{(L_2 + M_2 + 2)(L_2 + M_2 + 1)}{(2L_2 + 1)(2L_2 + 3)} \right)^{1/2} \delta_{M_1, M_2+1} & \text{if } \alpha = - \quad (\text{Eq. (S46)}) \end{cases} \tag{S87}$$

$$C_2^\alpha(L_1, M_1, M_2) = \begin{cases} \left( \frac{(L_1 + M_1 + 1)(L_1 - M_1 + 1)}{(2L_1 + 1)(2L_1 + 3)} \right)^{1/2} \delta_{M_1, M_2} & \text{if } \alpha = z \quad (\text{Eq. (S43)}) \\ -\left( \frac{(L_1 + M_1 + 2)(L_1 + M_1 + 1)}{(2L_1 + 1)(2L_1 + 3)} \right)^{1/2} \delta_{M_1+1, M_2} & \text{if } \alpha = + \quad (\text{Eq. (S51)}) \\ \left( \frac{(L_1 - M_1 + 2)(L_1 - M_1 + 1)}{(2L_1 + 1)(2L_1 + 3)} \right)^{1/2} \delta_{M_1, M_2+1} & \text{if } \alpha = - \quad (\text{Eq. (S47)}) \end{cases} \tag{S88}$$

where

$$F(K, L) = \sum_{m=0}^{K} \frac{4^m B_{mL}}{4^K (K-m)!(K-m)!(L+2m)!} \tag{S89}$$

$$B_{m,L} = \frac{8\pi(L+2m)!(L+m+1)! 2^L}{m!(2L+2m+2)!} \quad m, L \in \mathbb{N}_0 \tag{S90}$$

### 4.3.2 Case $L_1 + 1 = L_2$

$$\frac{\left\langle \phi_I^{(L_1,M_1)} \middle| \mathbf{\Omega}_\alpha \middle| \phi_J^{(L_2,M_2)} \right\rangle}{\left| \phi_I^{(L_1,M_1)} \right| \left| \phi_J^{(L_2,M_2)} \right|}$$

$$= \left( \frac{\det(2A_I) \det(2A_J)}{\det(A_{IJ}) \det(A_{IJ})} \right)^{3/4} \left( \frac{p_{II}}{q_I} \right)^{K_I} \left( \frac{p_{JJ}}{q_J} \right)^{K_J} \left( \frac{p_{IJ}}{\sqrt{q_I q_J}} \right)^{L_1} C_2^\alpha(L_1, M_1, M_2) \left( q_J F(K_I, L_1) F(K_J, L_2) \right)^{-1/2} \times$$

$$\left[ \frac{G_I^{\mathbf{\Omega}} p_{JJ}}{p_{IJ}} \sum_{m=1}^{\min(K_I, K_J+1)} \left( \frac{p_{IJ} p_{IJ}}{p_{II} p_{JJ}} \right)^m \left( \frac{B_{(m-1)L_2}}{(K_I - m)!(K_J - m + 1)!(2m + L_2 - 2)!} \right) \right.$$

$$\left. G_J^{\mathbf{\Omega}} \sum_{m=0}^{\min(K_I, K_J)} \left( \frac{p_{IJ} p_{IJ}}{p_{II} p_{JJ}} \right)^m \left( \frac{B_{mL_1}}{(K_I - m)!(K_J - m)!(2m + L_1)!} \right) \right]$$

$$= \left( \frac{\det(2A_I) \det(2A_J)}{\det(A_{IJ}) \det(A_{IJ})} \right)^{3/4} \left( \frac{p_{II}}{q_I} \right)^{K_I} \left( \frac{p_{JJ}}{q_J} \right)^{K_J} \left( \frac{p_{IJ}}{\sqrt{q_I q_J}} \right)^{L_1} C_2^\alpha(L_1, M_1, M_2) \left( q_J \right)^{-1/2} \times$$

$$\left[ \frac{G_I^{\mathbf{\Omega}} p_{JJ}}{p_{IJ}} \sum_{m=1}^{\min(K_I, K_J+1)} \left( \frac{p_{IJ} p_{IJ}}{p_{II} p_{JJ}} \right)^m H_2(m, K_J, K_I, L_2) + G_J^{\mathbf{\Omega}} \sum_{m=0}^{\min(K_I, K_J)} \left( \frac{p_{IJ} p_{IJ}}{p_{II} p_{JJ}} \right)^m H_1(m, K_I, K_J, L_1) \right]$$
$$\tag{S91}$$

with

$$p_{XY} = \frac{1}{2} \boldsymbol{v}_X \boldsymbol{A}_{IJ}^{-1} \boldsymbol{v}_Y \quad \text{with} \quad X, Y \in \{I, J\} \tag{S92}$$

$$q_X = \frac{1}{2} \boldsymbol{v}_X^{\mathrm{T}} \boldsymbol{A}_X^{-1} \boldsymbol{v}_X \quad \text{with} \quad X \in \{I, J\} \tag{S93}$$

$$H_1(m, K_1, K_2, L) = \frac{B_{mL}}{(K_1 - m)!(K_2 - m)!(2m + L)!} \left[ F(K_1, L+1) F(K_2, L) \right]^{-1/2} \tag{S94}$$

$$H_2(m, K_1, K_2, L) = \frac{B_{(m-1)L}}{(K_1 - m + 1)!(K_2 - m)!(2m + L - 2)!} \left[ F(K_1, L) F(K_2, L-1) \right]^{-1/2} \tag{S95}$$

$$G_I^{\boldsymbol{\Omega}} = \begin{cases} \sum_n \dfrac{q_n (\boldsymbol{A}_J \boldsymbol{A}_{IJ}^{-1} \boldsymbol{v}_I)_n}{m_n (E_i - E_f)} & \text{if } \boldsymbol{\Omega} = \boldsymbol{\mu}_{\mathrm{if}}^{(\mathrm{v})} \\ \sum_n q_n (\boldsymbol{A}_{IJ}^{-1} \boldsymbol{v}_I)_n & \text{if } \boldsymbol{\Omega} = \boldsymbol{\mu}_{\mathrm{if}}^{(\mathrm{l})} \end{cases}$$

$$G_J^{\boldsymbol{\Omega}} = \begin{cases} \sum_n \dfrac{q_n (\boldsymbol{A}_J \boldsymbol{A}_{IJ}^{-1} \boldsymbol{v}_J - \boldsymbol{v}_J)_n}{m_n (E_i - E_f)} & \text{if } \boldsymbol{\Omega} = \boldsymbol{\mu}_{\mathrm{if}}^{(\mathrm{v})} \\ \sum_n q_n (\boldsymbol{A}_{IJ}^{-1} \boldsymbol{v}_J)_n & \text{if } \boldsymbol{\Omega} = \boldsymbol{\mu}_{\mathrm{if}}^{(\mathrm{l})} \end{cases} \tag{S96}$$

$$C_1^\alpha(L_2, M_2, M_1) = \begin{cases} \left( \dfrac{(L_2 + M_2 + 1)(L_2 - M_2 + 1)}{(2L_2 + 1)(2L_2 + 3)} \right)^{1/2} \delta_{M_1, M_2} & \text{if } \alpha = z \quad (\text{Eq. (S42)}) \\ \left( \dfrac{(L_2 - M_2 + 2)(L_2 - M_2 + 1)}{(2L_2 + 1)(2L_2 + 3)} \right)^{1/2} \delta_{M_1 + 1, M_2} & \text{if } \alpha = + \quad (\text{Eq. (S50)}) \\ -\left( \dfrac{(L_2 + M_2 + 2)(L_2 + M_2 + 1)}{(2L_2 + 1)(2L_2 + 3)} \right)^{1/2} \delta_{M_1, M_2 + 1} & \text{if } \alpha = - \quad (\text{Eq. (S46)}) \end{cases} \tag{S97}$$

$$C_2^\alpha(L_1, M_1, M_2) = \begin{cases} \left( \dfrac{(L_1 + M_1 + 1)(L_1 - M_1 + 1)}{(2L_1 + 1)(2L_1 + 3)} \right)^{1/2} \delta_{M_1, M_2} & \text{if } \alpha = z \quad (\text{Eq. (S43)}) \\ -\left( \dfrac{(L_1 + M_1 + 2)(L_1 + M_1 + 1)}{(2L_1 + 1)(2L_1 + 3)} \right)^{1/2} \delta_{M_1 + 1, M_2} & \text{if } \alpha = + \quad (\text{Eq. (S51)}) \\ \left( \dfrac{(L_1 - M_1 + 2)(L_1 - M_1 + 1)}{(2L_1 + 1)(2L_1 + 3)} \right)^{1/2} \delta_{M_1, M_2 + 1} & \text{if } \alpha = - \quad (\text{Eq. (S47)}) \end{cases} \tag{S98}$$

where

$$F(K, L) = \sum_{m=0}^{K} \frac{4^m B_{mL}}{4^K (K - m)!(K - m)!(L + 2m)!} \tag{S99}$$

$$B_{m,L} = \frac{8\pi (L + 2m)!(L + m + 1)! 2^L}{m!(2L + 2m + 2)!} \quad m, L \in \mathbb{N}_0 \tag{S100}$$



\end{document}